%% file: main.tex
\documentclass[a4paper,11pt]{article}
\usepackage{jcappub} % for details on the use of the package, please see the JINST-author-manual
\usepackage{lineno}
\usepackage{orcidlink}
\usepackage{ulem}
\usepackage{stfloats}
\usepackage{multirow}
\usepackage{makecell}
\usepackage{array}

\usepackage[T1]{fontenc}
\usepackage{caption}
\usepackage{natbib}
\usepackage{graphicx}	% Including figure files

\usepackage{amsmath}	% Advanced maths commands
\usepackage{amssymb}	% Extra maths symbols
\usepackage{xcolor}

\newcommand{\bfx}{{\boldsymbol{x}}}
\newcommand{\bfk}{{\boldsymbol{k}}}
\newcommand{\bfn}{{\boldsymbol{n}}}

\newcommand{\X}{{e^{-\pi \mathrm{i}n_x}}}
\newcommand{\Y}{{e^{-\pi \mathrm{i}n_y}}}
\newcommand{\Z}{{e^{-\pi \mathrm{i}n_z}}}
\newcommand{\Ij}[1]{{e^{-\frac{\pi \mathrm{i}n_i}{2^{#1}}}}}

\newcommand{\Xj}[1]{{e^{-\frac{\pi \mathrm{i}n_x}{2^{#1}}}}}
\newcommand{\Yj}[1]{{e^{-\frac{\pi \mathrm{i}n_y}{2^{#1}}}}}
\newcommand{\Zj}[1]{{e^{-\frac{\pi \mathrm{i}n_z}{2^{#1}}}}}

\newcommand{\mpch}{h^{-1} {\rm Mpc}}
\newcommand{\mpcht}{h^{-3} {\rm Mpc^3}}
\newcommand{\gpch}{h^{-1} {\rm Gpc}}

\newcommand{\hmpc}{h {\rm Mpc}^{-1}}
\newcommand{\hmpct}{h^3 {\rm Mpc}^{-3}}

\newcommand{\sumjm}{{\sum_{j=1}^m}}

\newcommand{\prodjh}{{\prod_{j=1}^h}}
\newcommand{\prodjoh}{{\prod_{j=0}^h}}
\newcommand{\fsize}{7cm}

\newcommand{\interlace}{{\operatorname{int}}}

\title{\boldmath Accurate power spectrum estimation toward Nyquist limit}

\author[~\orcidlink{0009-0007-8965-3102}1,2]{Yipeng Wang,}
\author[~\orcidlink{0000-0002-9359-7170}1,2]{Yu Yu,}
\affiliation[1]{\textit{Department of Astronomy, Shanghai Jiao Tong University, 800 Dongchuan Road, Shanghai 200240, China}}
\affiliation[2]{\textit{Key Laboratory for Particle Astrophysics and Cosmology (MOE) and Shanghai Key Laboratory for Particle Physics and Cosmology, Shanghai 200240, China}}

\emailAdd{519070910011wyp@sjtu.edu.cn}
\emailAdd{yuyu22@sjtu.edu.cn}

\abstract{
The power spectrum, as a statistic in Fourier space, is commonly numerically calculated using the fast Fourier transform method to efficiently reduce the computational costs. 
To alleviate the systematic bias known as aliasing due to the insufficient sampling, the interlacing technique was proposed. 
We derive the analytical form of the shot noise under the interlacing technique,
which enables the exact separation of the Poisson shot noise from the signal in this case.
Thanks to the accurate shot noise subtraction, we demonstrate an enhancement in the accuracy of power spectrum estimation.
For the three dark matter samples with number density ranging from $10^{-2} \text{ Mpc}^{-3} h^3$ to $10^{-4} \text{ Mpc}^{-3} h^3$ analyzed on a coarse mesh of size $3\,\mpch$ with Cloud-in-Cell mass assignment scheme, the systematic bias is well under control up to the Nyquist frequency.
On the contrary, the bias induced by the estimator that ignore the aliasing on the shot noise typically exceeds the statistical uncertainty on the frequency beyond 0.85 times the Nyquist frequency. 
The good performance of our estimation allows an abatement in the computational cost by using low resolution and low order mass assignment scheme in the analysis for huge surveys and mocks.
}

\begin{document}
\maketitle
%\flushbottom

\onecolumn

\section{Introduction}
\label{sec:intro}
\input{introduction.tex}

\section{Formulae and estimators}
\label{sec:estimators}
\subsection{Power spectrum estimator}
\input{powerspectrum_estimator}

\subsection{No interlacing}
\input{powerspectrum_estimator_nointerlacing}

\subsection{Interlacing technique}
\input{powerspectrum_estimator_interlacing}

%\subsection{Error estimation}
%\input{error_estimation}

\section{Numerical tests}
\label{sec:results}
\input{numerical_test}

\subsection{Full dark matter sample}
\label{sec:dmsample}
\input{dmfull}
\subsection{Down-sampled dark matter}
\label{sec:dmlowsample}
\input{dm100pc}

\input{dm10pc}
\input{dm1pc}

\subsection{Halo sample}
\label{sec:halosample}
\input{halo_mass_order_10percent}

\input{halo_mass_order_1percent}

\section{Conclusion}
\label{sec:conclusion}
\input{conclusion.tex}

\acknowledgments
We thank Zhao Chen for the useful discussions and the anonymous referee for the valuable suggestions.
This work is supported by 
the National Key R\&D Program of China (2023YFA1607800, 2023YFA1607802), 
the National Science Foundation of China (grant numbers 12273020), 
the China Manned Space Project with numbers CMS-CSST-2021-A03, 
the ‘111’ Project of the Ministry of Education under grant number B20019, 
and the sponsorship from Yangyang Development Fund. 
This work made use of the Gravity Supercomputer at the Department of Astronomy, Shanghai Jiao Tong University.

\appendix
\section{Isotropic approximation for shot noise}
\input{shotnoise_expansion}

\section{Shot noise for high order interlacing}
\label{sec:shotnoise_highorder}
\input{shotnoise_higherorder.tex}

\bibliographystyle{JHEP}
\bibliography{citation.bib}

\label{lastpage}
\end{document}

%% file: introduction.tex
The power spectrum, usually denoted as $P(k)$ for three dimensional space, is an important statistical quantity in cosmological research. It is the Fourier transform counterpart of the two-point correlation function and reflects the magnitude of the matter density fluctuations at different spatial scales. 
The most accurate power spectrum estimation is the direct summation \cite{Sefusatti16}, but it is very time consuming as the computational cost is proportional to the number of grid points in $\bfk$ space times the number of galaxies.
Thus, the power spectrum is usually calculated using the Fast Fourier Transform (FFT) method.
By factorizing the calculation to even and odd series recursively, FFT reduces the computational complexity from $\mathcal{O}(N^2)$ to $\mathcal{O}(N \log N)$, with $N$ being the total number of grid points.
\par
Under the FFT method, the approximations made introduce a series of systematic effects that need to be corrected. The FFT method requires us to sample the original catalogue at specific grid points for subsequent computations. 
This sampling process is known as the mass assignment scheme. 
On one hand, reconstructing a continuous signal from discrete samples requires an interpolation scheme to preserve the sampling, and the choice of interpolation kernel is related to the mass assignment scheme used. 
The deviation caused by the variation in the mass assignment function is known as the window function effect, which needs to be corrected by the deconvolution. 
On the other hand, representing a continuous mass distribution 
by a discrete mass distribution introduces inherent background noise known as shot noise effect. 
On top of the two, the maximum frequency we can sample is determined by the frequency corresponding to the separation of grid points, and any signal with frequencies higher than this cannot be measured. 
According to the Shannon sampling theorem, in order to faithfully recover the original signal without loss or distortion, the sampling frequency must be higher than twice the maximum frequency component of the original signal, which requires a truncation of the signal at high frequencies. 
However, when analyzing large-scale structures, the frequencies of cosmological perturbations typically exceed the Nyquist frequency in the analysis, 
%making it impossible to satisfy the sampling theorem. 
%Therefore, the Nyquist frequency is usually lower than the frequency of the smallest scale in the signal, 
causing high-frequency components to be erroneously folded into the part within the sampling frequency range. 
This systematic error caused by the finite sampling frequency is known as aliasing. 
Among these systematic errors, the window function effect and shot noise effect are independent of the distribution of the sample and can be precisely solved and eliminated.  The analytical expressions for the window function term and shot noise correction term was provided in \cite{Jing05}. 
On the contrary, the bias resulting from aliasing is related to the signal itself, and is difficult to remove without prior knowledge.
\par
To mitigate the contamination of the signal by aliasing, one approach is to apply a low-pass filter to the signal before Fourier transformation. However, as a interpolation kernel, the computational cost of a low-pass filter increases with the rigor of the filter, making the advantage of using the FFT method for optimization and acceleration over direct summation less significant. In cosmological research, there have been various attempts to mitigate the effects of aliasing on power spectrum measurements. 
 An iterative elimination method by assuming that the real power spectrum near Nyquist frequency can be approximated as a power law was proposed in \cite{Jing05}.
%\citet{Jing05} proposed a mathematical expression for aliasing effects by assuming that the power spectrum near the Nyquist frequency can be approximated by a power-law form. He also provided a scheme based on an iterative process to reduce the impact of aliasing.
For selecting the ideal window function to minimize sampling effects certain criteria was proposed and the Daubechies wavelet transform-based window function is suggested as a candidate for minimizing sampling effects \cite{Cui08}.
Based on the Daubechies 6 wavelet basis and third-order B-splines a mass assignment scheme was further proposed in \cite{Yang09}.
Interlacing technique was proposed by \cite{CHEN74}, and later introduced in the $AP^3M$ N-body code \cite{Couchman91}. Originally used in television signals or image processing, interlacing is a method to enhance image or signal quality by effectively reducing aliasing effects using two interlaced grids with only twice the computational cost of standard operations,
 the reduction in systematic errors achieved by this method was quantified in \cite{Sefusatti16}.
\par
In this paper, we focus on the precise derivation and elimination of the shot noise term under the interlacing technique, further improving the accuracy of the power spectrum  estimation.
The paper is organized as follows. 
Chapter \ref{sec:estimators} begins with an derivation of the analytical relationship between the power spectrum measurement from FFT and real signals. 
It also reviews common forms of power spectrum estimation in the literature and the assumptions underlying these estimation methods. 
In the third part of Chapter \ref{sec:estimators}, the interlacing technique is introduced, explaining its principles for reducing aliasing effects. 
The precise form of the correct shot noise term under this technique is derived, highlighting its significance in capturing aliasing effects in the shot noise signal. 
%Recommended estimation methods using this technique are presented. Also, upper bounds of the residual error for different estimators are derived. 
%These residual error estimations serve to validate the shot noise correction term derived under the interlacing technique and demonstrate the potential for improved precision with our recommended estimators, particularly when the shot noise component is significant compared to high-frequency signal components.
%\par
Numerical tests are performed in Chapter \ref{sec:results} to quantify the obtained power spectrum using the various estimation methods discussed. 
%The results are compared with high-resolution Fast Fourier Transform results, demonstrating that the residual errors for different estimators are negligible within the precision range\red{???}. 
The simulations provide evidence of the precision advantage of one of the estimators, especially at small scales, near the Nyquist frequency, and with lower-order window function sampling and lower sample particle density. 
This advantage is particularly important in future large-scale data analysis, as the sky coverage and volume occupation is dramatically increased so that computational efficiency is crucial.
We summarize the results in Chapter \ref{sec:conclusion}.

%% file: powerspectrum_estimator.tex
We focus on a simulated halo catalogue within a periodical cubic box of volume $V=L^3$. $L$ represents the physical/comoving size of the box.
%This data can also be viewed as an infinite distribution with periodicity, a hypothesis that is generally applicable to a universe consistent with cosmological principles. 
For the power spectrum calculation,
first we need to use a mass assignment scheme to convert a catalogue to a field.
%\red{use full name plus short name in brackets as the first time.}
Usually, Nearest Grid Point (NGP), Cloud-In-Cell (CIC), Triangular Shaped Cloud (TSC) and Piecewise Cubic Spline (PCS) scheme is adopted.
We refer the readers to \cite{Sefusatti16} for the real space kernels of the above mass assignments.
We express the halo number density $n(\bfx)$ in a dimensionless form
\begin{equation}
    \delta^f(\bfx) = \frac{n^f(\bfx)}{\bar{n}}-1\ ,
\end{equation}
where $\bar{n}$ is the mean density within the volume $V$, and the superscript $f$ represents that the density field is sampled using one of the above mass assignment scheme. 
Under the condition of periodicity, we Fourier transform  the overdensity field,
\begin{equation}
    \delta^f(\bfk) = \frac{1}{V}\int_{V}{\delta^f(\bfx)e^{i\bfk\cdot\bfx}d\bfx\ .}
\end{equation}
%\par
In Fourier space, the mass assignment schemes induce window effect on the resulting field,
\begin{equation}
W(\bfk)=\left[
\frac{\sin(\pi k_1/2k_N)}{\pi k_1/2k_N}
\frac{\sin(\pi k_2/2k_N)}{\pi k_2/2k_N}
\frac{\sin(\pi k_3/2k_N)}{\pi k_3/2k_N}
\right]^p\ .
\label{eqn:window}
\end{equation}
For NGP, CIC, TSC and PSC, $p=1,2,3$ and $4$, respectively.
$k_N = \frac{\pi N_g}{L}$ represents the Nyquist frequency, and $N_g$ represents the number of grid points in one dimension. 
%For the calculation of the power spectrum of a discrete mass field using the Fast Fourier Transform\brown{\sout{, the following equation \citep{Jing05} is used}}:
Following the derivation, conventions and expressions in \cite{Jing05}, the resulting power spectrum is 
\begin{equation}
\begin{aligned}
\langle|\delta^f(\bfk)|^2\rangle
%&=\sum_\bfn|W(\bfk+2k_N\bfn)|^2 \left[ P(\bfk+2k_N\bfn)+\frac{1}{N} \right] \\
&=\sum_\bfn|W(\bfk+2k_N\bfn)|^2P(\bfk+2k_N\bfn)+\frac{1}{N}\sum_\bfn|W(\bfk+2k_N\bfn)|^2\ .
\end{aligned}
\label{eqn:allterm}
\end{equation}
%\par
We refer the first term in the \text{r.h.s.} of Eq.\,(\ref{eqn:allterm}) as the signal term. 
It is the summation of the power spectrum with window effect over a series of shifted images,
which is known as {\it aliasing}.
The $\bfn\neq 0$ components represent the contribution of aliased images, which boosts the power near Nyquist frequency if the sampling rate is not sufficiently high.
The last term in Eq.\,(\ref{eqn:allterm}) is the shot noise contribution.
Note that the shot noise $1/N$ also suffers from the window function and aliasing effect.
%However, the summation over images means that the obtained power is usually overestimated, which is known as {\it aliasing}.

%% file: powerspectrum_estimator_nointerlacing.tex
As in \cite{Jing05}, we denote the shot noise term as 
$\frac{1}{N}C(\bfk)\equiv\frac{1}{N}\sum_\bfn|W(\bfk+2k_N\bfn)|^2$.
%\begin{equation}
%\frac{1}{N}C(\bfk)\equiv\frac{1}{N}\sum_\bfn|W(\bfk+2k_N\bfn)|^2\ ,
%\label{eqn:sumw}
%\end{equation}
From the identity
\begin{equation}
\begin{aligned}
\sum_n \frac{1}{(x-\pi n)^{q}}=\frac{(-1)^{q-1}}{{(q-1)}!}\frac{d^{q-1}}{dx^{q-1}}\cot x\ ,
\end{aligned}
\label{eqn:mathidentity}
\end{equation}
the summation over infinite images of $W$ in one dimension can be derived.
Let $q=2p$ and $x=\pi k/2k_N$,
\begin{equation}
\begin{aligned}
\sum_n|W(k+2k_Nn)|^2&=\sum_n\left(\frac{\sin(\pi k/2k_N+n\pi)}{\pi k/2k_N+n\pi}\right)^{2p}
%\\=\sin^{2p}(\pi k/2k_N)\sum_n \frac{1}{(\pi k/2k_N+n\pi)^{2p}}
\\&=\sin^{2p}(x)\sum_n \frac{1}{(x+n\pi)^{2p}}
\\&={\sin^{q}x}\frac{(-1)^{q-1}}{{(q-1)}!}\frac{d^{q-1}}{dx^{q-1}}\cot x\ .
\end{aligned}
\label{eqn:sumw1d}
\end{equation}
We denote the result as $S_q[\sin^2x]$ since it is a polynomial function of $\sin^2x$ up to order of $p-1$,
\begin{equation}
\begin{aligned}
%%\\ &\equiv S_q[\sin^2x]
S_q[\sin^2x]
&=\left\{
\begin{aligned}
&1\ , & \text{NGP}\ ,\\
&1-\frac{2}{3}\sin^2x\ , & \text{CIC}\ ,\\
&1-\sin^2x+\frac{2}{15}\sin^4x\ , & \text{TSC}\ ,\\
&1-\frac{4}{3}\sin^2x+\frac{2}{5}\sin^4x-\frac{4}{315}\sin^6x\ , & \text{PSC}\ .\\
\end{aligned}
\right.
\end{aligned}
\label{eqn:sqpoly}
\end{equation}
%Thus, 
%\begin{equation}
%C(\bfk)=\prod_i S_{2p}\left[\sin^2\left(\frac{\pi k_i}{2k_N}\right)\right]
%\label{eqn:shotnoise2}
%\end{equation}
Thus, $C(\bfk)$ has the exact analytical expression 
\begin{equation}
C(\bfk)=\left\{
\begin{aligned}
&1\ , & \text{NGP}\ ,\\
&\prod_i\left[1-\frac{2}{3}\sin^2\left(\frac{\pi k_i}{2k_N}\right)\right]\ , & \text{CIC}\ ,\\
&\prod_i\left[1-\sin^2\left(\frac{\pi k_i}{2k_N}\right)+\frac{2}{15}\sin^4\left(\frac{\pi k_i}{2k_N}\right)\right]\ , & \text{TSC}\ ,\\
&\prod_i\left[1-\frac{4}{3}\sin^2\left(\frac{\pi k_i}{2k_N}\right)+\frac{2}{5}\sin^4\left(\frac{\pi k_i}{2k_N}\right)-\frac{4}{315}\sin^6\left(\frac{\pi k_i}{2k_N}\right)\right]\ , & \text{PSC}\ .\\
\end{aligned}
\right.
\label{eqn:shotnoise}
\end{equation}
We emphasize that the shot noise term can be precisely subtracted from Eq.\,(\ref{eqn:allterm}).
The isotropic versions of the shot noise term $C(k)$ except for the PSC scheme are provided in \cite{Jing05}, and we also list the isotropic versions in the Appendix.
\par
Neglecting the contamination from aliased images on the signal, i.e. assuming that 
$\sum_\bfn|W(\bfk+2k_N\bfn)|^2P(\bfk+2k_N\bfn)\approx P(\bfk)|W(\bfk)|^2$,
a straightforward estimator is 
\begin{equation}
\label{eqn:noint-shot-win}
\hat{P}(\bfk)=\frac{\langle|\delta^f(\bfk)|^2\rangle-\frac{1}{N}C(\bfk)}{|W(\bfk)|^2}\ .
\end{equation}
In this estimator proposed in \cite{Jing05}, the shot noise term is subtracted precisely before correcting the window function in the signal term.
We expect that the discarded summation leads to the aliasing effect, a boost of the power spectrum amplitude near the Nyquist frequency.
%\purple{It is a boost isn't it? Since we take de facto $\sum_\bfn|W(\bfk+2k_N\bfn)|^2P(\bfk+2k_N\bfn)$ over $|W(\bfk)|^2$ as $\hat{P}(\bfk)$}

Under a different approximation 
$\sum_\bfn|W(\bfk+2k_N\bfn)|^2P(\bfk+2k_N\bfn)\approx P(\bfk)\sum_\bfn|W(\bfk+2k_N\bfn)|^2= P(\bfk)C(\bfk)$,
and thus $\langle|\delta^f(\bfk)|^2\rangle\approx [P(\bfk)+\frac{1}{N}]C(\bfk)$,
the corresponding estimator is
\begin{equation}
\label{eqn:noint-win-shot}
\hat{P}(\bfk)=\frac{\langle|\delta^f(\bfk)|^2\rangle}{C(\bfk)}-\frac{1}{N}\ .
\end{equation}
For this estimator, the window effect is corrected first and the $1/N$ shot noise is subtracted in the end.
We caution the readers that the window effect here is $C(\bfk)$ instead of $W^2(\bfk)$.
This correction can be performed on the field level, by dividing the effective window function $\sqrt{C(\bfk)}$ from the Fourier transformed field $\delta(\bfk)$.
This is also the approach adopted in \texttt{Nbodykit}\footnote{\url{https://github.com/bccp/nbodykit}, commit 376c9d7
}~\cite{Hand18} for the power spectrum estimation without interlacing technique.

%% file: powerspectrum_estimator_interlacing.tex
\begin{figure*}[htbp]
\centerline{\includegraphics[width=\fsize]{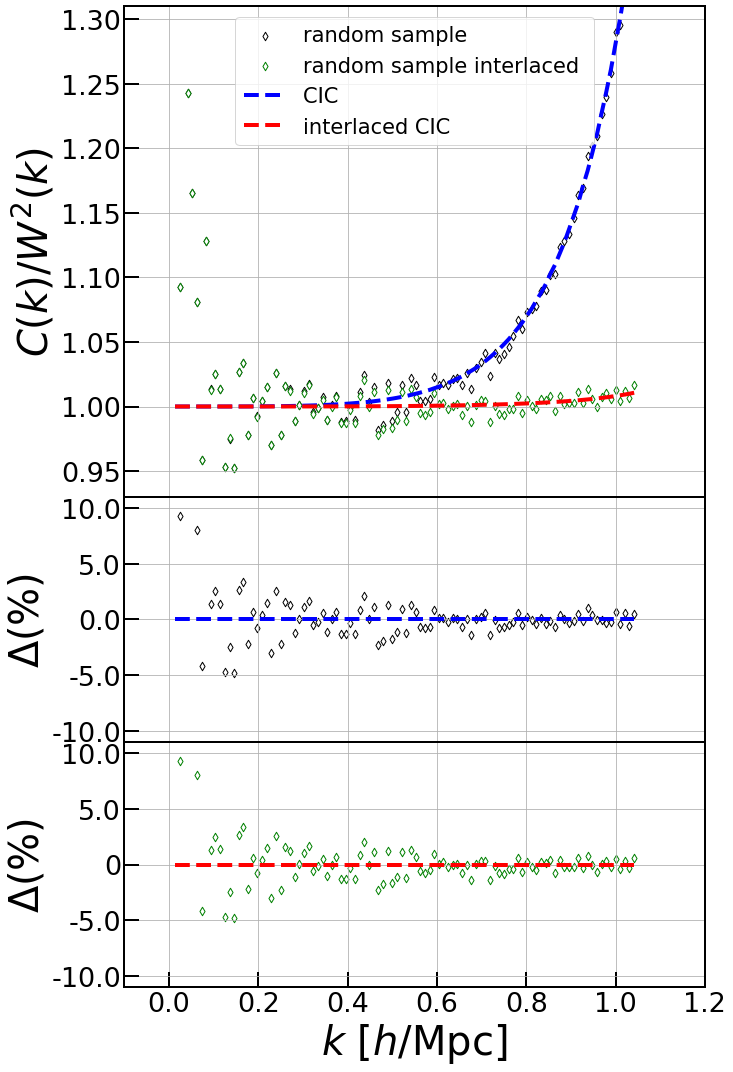}
\includegraphics[width=\fsize]{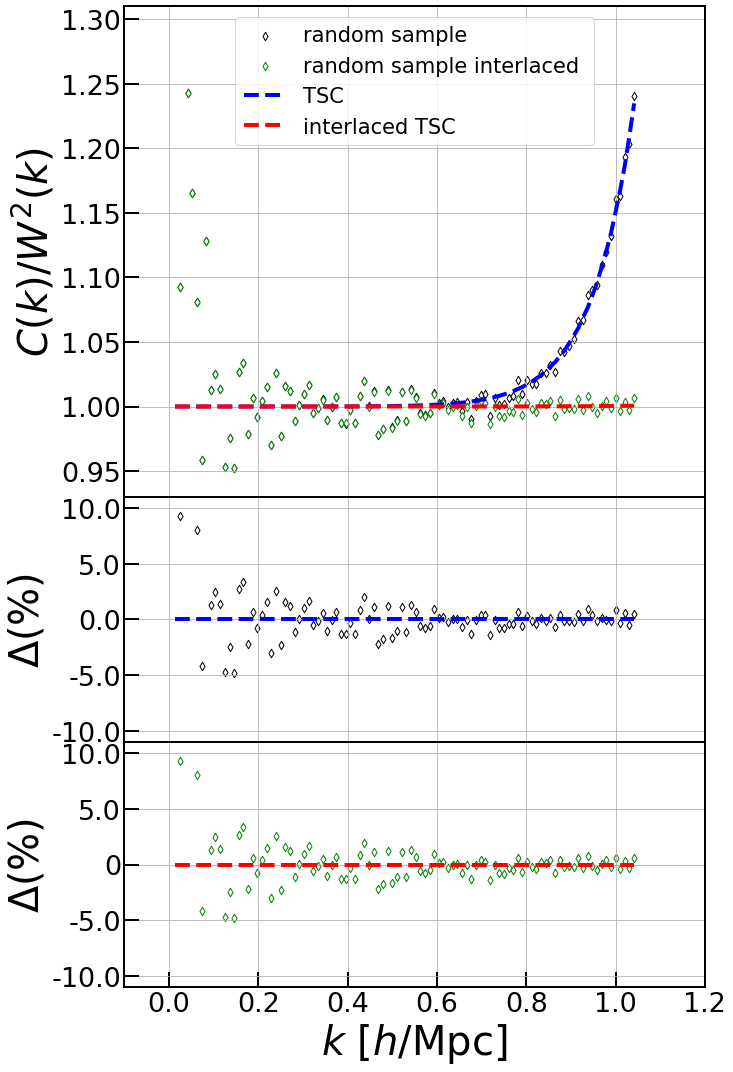}}
\caption{The comparison of the analytical shot noise expressions with the measurement from Poisson-distributed random samples. 
On the left side are the results for the CIC window function, and on the right side are the results for the TSC window function. 
In the top panels the data points represent the $\langle\delta^f(k)^2\rangle/W^2(k)$ obtained from the random sample, the blue dashed lines follow Eq.\,(\ref{eqn:shotnoise}) and the red lines follow Eq.\,(\ref{eqn:interlacedshotnoise_brief}).
The middle panels show the results scaled to the Eq.\,(\ref{eqn:shotnoise}) and to the Eq.\,(\ref{eqn:interlacedshotnoise_brief}) for the bottom panels.
The analytical expressions well describe the shot noise with interlacing technique.
}
\label{fig:shot_show}
\end{figure*}

\cite{hockney21} proposed the interlacing technique, which was lately applied to the $AP^3M$ N-body code~\cite{Couchman91}. 
This technique samples the signals twice with the second mesh having an overall half-cell offset $(H/2,H/2,H/2)$ relative to the primary mesh.
The odd terms in the aliasing images will be eliminated by taking the arithmetic average of the two samplings. As such, the effects of aliasing can be reduced, leading to an improved estimation of the power spectrum.
We denote the average of the two samplings as $\tilde\delta^f(\bfk)$, and
mathematically,
\begin{equation}
\begin{aligned}
\tilde\delta^f(\bfk)&=\frac{1}{2}\left[\delta^f_1(\bfk)+\delta^f_2(\bfk)\right]
\\&=\frac{1}{2}\left[\delta^f_1(\bfk)+\delta^f_2(\bfk')e^{-\mathrm{i} \bfk'\cdot \boldsymbol{H}/2}\right]\\
&=\sum_\bfn{\theta_\bfn{\delta}^f(\bfk+2k_N\bfn)}\ .
\end{aligned}
\label{eqn:interlace}
\end{equation}
Here, $\theta_\bfn$ constitutes an equivalent expression of the interlacing scheme in the algebraic space, defined as

\begin{equation}
\begin{aligned}
\theta_\bfn &\equiv \frac12 (1+(-1)^{n_x+n_y+n_z}) \\
&=\left\{
\begin{array}{ll}
1\ , & n_x+n_y+n_z \text{ is even}\ ,\\
0\ , & n_x+n_y+n_z \text{ is odd}\ .\\
\end{array}\right.
\end{aligned}
\label{eqn:interlace_index}
\end{equation}
Again, $\bfn = 0$ is the true signal term.
In most cases and also true for cosmological matter power spectrum, the signal decreases with increasing frequency,
and thus the most dominant contribution is from $|\bfn|= 1$.
With the interlacing technique above, the aliasing effect from all odd pixels is eliminated.
This approach smooths the sampling near the Nyquist frequency. 
At the same time, it should be noted that the shot noise is also affected by this smoothing effect. 
%Under no-interlacing circumstances, the shot noise term Eq. (\ref{eqn:shotnoise}) is derived from the identity,
%\begin{equation}
%\begin{aligned}
%C(\bfk) &= \sum_\bfn|W(\bfk+2k_N\bfn)|^2\\
%&=\sum_{n_i}\sum_{n_j}\sum_{n_m}\left(\frac{\sin(\pi k_i/2k_N+\pi n_i)}{\pi k_i/2k_N+\pi n_i}
%\frac{\sin(\pi k_j/2k_N+\pi n_j)}{\pi k_j/2k_N+\pi n_j}
%\frac{\sin(\pi k_m/2k_N+\pi n_m)}{\pi k_m/2k_N+\pi n_m}\right)^{2p}\\
%&=S_{q}\left[\sin^2\left(\frac{\pi k_i}{2k_N}\right)\right]S_{q}\left[\sin^2\left(\frac{\pi k_j}{2k_N}\right)\right]S_{q}\left[\sin^2\left(\frac{\pi k_m}{2k_N}\right)\right]\ .
%\end{aligned}
%\label{eqn:shotnoise2identity}
%\end{equation}
%Here, $q=2p$, and $S_{q}$'s are polynomial functions of $\sin^2x$ up to order of $p-1$,
%\begin{equation}
%\begin{aligned}
%S_{q}[\sin^2x]&=\sum_n\left(\frac{\sin (x-\pi n)}{x-\pi n}\right)^q\\
%&={\sin^{q}x}\sum_n \frac{1}{(x-\pi n)^{q}}\\
%&={\sin^{q}x}\frac{(-1)^{q-1}}{{(q-1)}!}\frac{d^{q-1}}{dx^{q-1}}\cot x\ .
%\end{aligned}
%\label{eqn:identity}
%\end{equation}

The shot noise term with interlacing is 
$C^{\interlace}(\bfk)=\sum_{\bfn}\theta_\bfn|W(\bfk+2k_N\bfn)|^2$.
In other words, we are now interested in the summation of $\sum_\bfn|W(\bfk+2k_N\bfn)|^2$ but over even $n_x+n_y+n_z$ terms.
These terms can be split into two groups: all the $n_i$'s are even, or two of the three $n_i$'s are odd.
Let $n=2m$, the summation over   even terms in Eq.\,(\ref{eqn:sumw1d})  is reduced to 
\begin{equation}
\begin{aligned}
C_e(x) &\equiv \sum_{n \text{ is even}} \frac{\sin^{2p}(x+\pi n)}{(x+\pi n)^{2p}}
=\sum_{m} \frac{\sin^{2p}(x+2\pi m)}{(x+2\pi m)^{2p}}
=\frac{1}{2^{2p}}\frac{\sin^{2p}(x)}{\sin^{2p}(x/2)}S_{2p}[\sin^2(x/2)]\\
&=\cos^{q}(x/2)S_{q}[\sin^2(x/2)]\\
%% &=\left(\frac{1+\cos x}{2}\right)^{p}S_{p}\left[\left(\frac{1-\cos x}{2}\right)\right]\ 
&=\left\{
\begin{aligned}
&\cos^2\left(\frac{x}{2}\right)\ , &\text{NGP}\ ,\\
&\cos^4\left(\frac{x}{2}\right)\left[1-\frac{2}{3}\sin^2\left(\frac{x}{2}\right)\right]\ , &\text{CIC}\ ,\\
&\cos^6\left(\frac{x}{2}\right)\left[1-\sin^2\left(\frac{x}{2}\right)+\frac{2}{15}\sin^4\left(\frac{x}{2}\right)\right]\ , &\text{TSC}\ ,\\
&\cos^8\left(\frac{x}{2}\right)\left[1-\frac{4}{3}\sin^2\left(\frac{x}{2}\right)+\frac{2}{5}\sin^4\left(\frac{x}{2}\right)-\frac{4}{315}\sin^6\left(\frac{x}{2}\right)\right]\ , &\text{PSC}\ .\\
\end{aligned}
\right.
\end{aligned}
\end{equation}
Here, $x = \frac{\pi k}{2k_N}$. A similar derivation of the above relation is shown in \cite{Sefusatti16} to obtain the relation between the residuals of the power spectrum estimator with and without interlacing.
The summation over odd terms can be derived by letting $n=2m+1$, 
\begin{equation}
\begin{aligned}
C_o(x) &\equiv \sum_{n \text{ is odd}} \frac{\sin^{2p}(x+\pi n)}{(x+\pi n)^{2p}}\\
&=\sum_m \frac{\sin^{2p}(x+\pi+2\pi m)}{(x+\pi+2\pi m)^{2p}}
=\frac{1}{2^{2p}}\frac{\sin^{2p}(x+\pi)}{\sin^{2p}(\frac{x+\pi}{2})}S_{2p}[\sin^2(\frac{x+\pi}{2})]\\
&=\cos^{2p}\left(\frac{x+\pi}{2}\right)S_{2p}\left[\sin^2\left(\frac{x+\pi}{2}\right)\right]
=\sin^{q}(x/2)S_{q}[\cos^2(x/2)]\\ 
&=\left\{
\begin{aligned}
&\sin^2\left(\frac{x}{2}\right)\ , &\text{NGP}\ ,\\
&\sin^4\left(\frac{x}{2}\right)\left[1-\frac{2}{3}\cos^2\left(\frac{x}{2}\right)\right]\ , &\text{CIC}\ ,\\
&\sin^6\left(\frac{x}{2}\right)\left[1-\cos^2\left(\frac{x}{2}\right)+\frac{2}{15}\cos^4\left(\frac{x}{2}\right)\right]\ , &\text{TSC}\ ,\\
&\sin^8\left(\frac{x}{2}\right)\left[1-\frac{4}{3}\cos^2\left(\frac{x}{2}\right)+\frac{2}{5}\cos^4\left(\frac{x}{2}\right)-\frac{4}{315}\cos^6\left(\frac{x}{2}\right)\right]\ , &\text{PSC}\ .\\
\end{aligned}
\right.
\end{aligned}
\end{equation}
Thus, the exact shot noise term with interlacing is 
\begin{equation}
%\begin{aligned}
C^{\interlace}(\bfk)
%&=\sum_{\bfn}\theta_\bfn|W(\bfk+2k_N\bfn)|^2\\
=\prod_i C_e\left(\frac{\pi k_i}{2k_N}\right)+\sum_{(i,j,m)\in A_3} C_e\left(\frac{\pi k_i}{2k_N}\right)C_o\left(\frac{\pi k_j}{2k_N}\right)C_o\left(\frac{\pi k_m}{2k_N}\right)\ .
%\end{aligned}
\label{eqn:interlacedshotnoise_brief}
\end{equation}
Here, the summation in the second term is over $A_3$, the alternating group of degree $3$, which collects the contributions from $(i,j,m)=(1,2,3),(2,3,1)$ and $(3,1,2)$.

To exam the accuracy of the above shot noise formula, we calculate the power spectrum of a random sample with number density of $n_s = 1.12\times 10^{-3} \text{ Mpc}^{-3} h^3$.
We show the performance of the Eq.\,(\ref{eqn:interlacedshotnoise_brief}) in Figure~\ref{fig:shot_show}, along with the result for Eq.\,(\ref{eqn:shotnoise})  (also appeared in \cite{Jing05}).
The results are presented in the form of $C(k)/W^2(k)$ to line up with the second term, i.e. the assumed shot noise contribution, in estimator Eq.\,(\ref{eqn:noint-shot-win}).
The analytical expressions describe the data points from the numerical calculation well, with the residual scattering around the expectation.
The accuracy of the interlaced shot noise formula 
lines up with the accuracy of the case without interlacing.

The first term in Eq.\,(\ref{eqn:interlacedshotnoise_brief}) is usually larger than the second term which includes sine functions from two odd terms.
This very expression of the leading term is similar to Eq.\,(\ref{eqn:sqpoly}), with $\sin^2x$ being replaced by $\sin^2(x/2)$ and an extra cosine prefactors $\cos^{2p}(x/2)$ induced by interlacing.
Notice that for NGP mass assignment without interlacing, the aliasing effect exactly compensates the window effect, leading to exactly $1/N$ shot noise power.
However, when interlacing is adopted, the shot noise for NGP is lower than $1/N$ at small scale.
Naively using $1/N$ as the shot noise like Eq.\,(\ref{eqn:noint-shot-win}) leads to an over-subtraction of the shot noise.
\par
Under the same approximation, $\sum_\bfn|W(\bfk+2k_N\bfn)|^2P(\bfk+2k_N\bfn)\approx P(\bfk)|W(\bfk)|^2$ in deriving the estimator Eq.\,(\ref{eqn:noint-shot-win}), 
the new estimator for interlacing is 
\begin{equation}
\label{eqn:int-shot-win}
\hat{P}(\bfk)=\frac{\langle|\tilde\delta^f(\bfk)|^2\rangle-\frac{1}{N}C^{\interlace}(\bfk)}{|W(\bfk)|^2}\ .
\end{equation}
Again, in this estimator the shot noise term is subtracted precisely before correcting the window function in the signal term.
Since aliasing is significantly reduced by the interlacing technique,
an accurate shot noise subtraction is essential to avoid the systematic bias.

The interlacing technique reduces aliasing both in the signal term and in the shot noise term.
Under this consideration, we can directly discard the summation as
$\sum_\bfn|W(\bfk+2k_N\bfn)|^2\left[P(\bfk+2k_N\bfn)+\frac{1}{N}\right]\approx [P(\bfk)+\frac{1}{N}]|W(\bfk)|^2$,
and thus the corresponding estimator is
\begin{equation}
\label{eqn:int-win-shot}
\hat{P}(\bfk)=\frac{\langle|\tilde\delta^f(\bfk)|^2\rangle}{|W(\bfk)|^2}-\frac{1}{N}\ .
\end{equation}
In this estimator, the shot noise can be easily separated as a constant term throughout the entire computation process, which is also the power spectrum calculation approach in \texttt{Nbodykit}~\cite{Hand18} for interlacing. 
Comparing this estimation with Eq.\,(\ref{eqn:int-shot-win}), it can be observed that the shot noise assumed here is $\frac{1}{N} W^2(\bfk)$ instead of $\frac{1}{N} C^{\interlace}(\mathbf{k})$.  
Although the difference between these two assumed shot noise forms is small, the approach in Eq.\,(\ref{eqn:int-win-shot}) still underestimates the contribution of shot noise, leading to an overestimation of the power spectrum. 
We summarize the above four estimators in Table~\ref{tab:estimators}.
\par

Note that the interlacing technique here refers to use 2 interlaced grids, which eliminates the main contamination from $|\bfn|=1$.
To eliminate more images, we have to use more interlaced grids.
In Appendix~\ref{sec:shotnoise_highorder}, we also provide the analytical shot noise expressions with multiple interlaced grids  and various mass assignment schemes.

\begin{table*}
\centering
\caption{
The summary of the assumptions and properties of various power spectrum estimators.
The first two rows do not involve interlacing and the last two adopt interlacing.
For the estimators in the first and third row, the shot noise is precisely subtracted first, then the window effect is deconvolved.
For the estimators in the second and fourth row, the window effect is deconvolved at field level,
then the shot noise is subtracted from the obtained power spectrum.
\label{tab:estimators}
}
\label{tab:example_table}

\begin{tabular}{l|l|l}% four columns, alignment for each
%\begin{tabular}{l|l|l}% four columns, alignment for each
\hline
Approximation & Estimator & Description\\
\hline
%% line1
\makecell*[l]{$\sum_\bfn|W(\bfk+2k_N\bfn)|^2P(\bfk+2k_N\bfn)$\\
 $\approx P(\bfk)|W(\bfk)|^2$} &
\makecell*[l]{$\hat{P}(\bfk)=\frac{\langle|\delta^f(\bfk)|^2\rangle-\frac{1}{N}C(\bfk)}{|W(\bfk)|^2}$, \\
Eq.\,(\ref{eqn:noint-shot-win})} &
\makecell*[{{p{3.5cm}}}]{Exact shot noise subtraction without interlacing.}\\
\hline
%% line2
\makecell*[l]{$\sum_\bfn|W(\bfk+2k_N\bfn)|^2P(\bfk+2k_N\bfn)$\\
$\approx P(\bfk)\sum_\bfn|W(\bfk+2k_N\bfn)|^2$} &
\makecell*[l]{$\hat{P}(\bfk)=\frac{\langle|\delta^f(\bfk)|^2\rangle}{C(\bfk)}-\frac{1}{N}$,\\
Eq.\,(\ref{eqn:noint-win-shot}) }&
\makecell*[{{p{3.5cm}}}]{Window effect $\sqrt{C(\bfk)}$ can be deconvolved at field level.}
\\
\hline
%% line3
\makecell*[l]{$\sum_\bfn|W(\bfk+2k_N\bfn)|^2P(\bfk+2k_N\bfn)$\\
$\approx P(\bfk)|W(\bfk)|^2$} &
\makecell*[l]{$\hat{P}(\bfk)=\frac{\langle|\tilde\delta^f(\bfk)|^2\rangle-\frac{1}{N}C^{\interlace}(\bfk)}{|W(\bfk)|^2}$,\\
Eq.\,(\ref{eqn:int-shot-win}) }&
\makecell*[{{p{3.5cm}}}]{Exact shot noise subtraction for interlacing.}
\\
\hline
%% line4
\makecell*[l]{$\sum_\bfn|W(\bfk+2k_N\bfn)|^2\left[P(\bfk+2k_N\bfn)+\frac{1}{N}\right]$\\
$\approx \left[P(\bfk)+\frac{1}{N}\right]|W(\bfk)|^2$} &
\makecell*[l]{$\hat{P}(\bfk)=\frac{\langle|\tilde\delta^f(\bfk)|^2\rangle}{|W(\bfk)|^2}-\frac{1}{N}$, \\
Eq.\,(\ref{eqn:int-win-shot}) }&
\makecell*[{{p{3.5cm}}}]{Window effect $|W(\bfk)|$ can be deconvolved at field level.}
\\
\hline
\end{tabular}
\end{table*}

%% file: numerical_test.tex
We conduct the numerical tests on both the dark matter particle samples and the halo samples simulated by  \texttt{Gadget-4} code \cite{GADGET4}.
The simulation box length is $L = 600\,h^{-1}\mathrm{Mpc}$.
%\sout{Gadget (Galaxies with Dark matter and Gas intEracT) is a code developed by Volker Springel at the Max Planck Institute for Astrophysics in Germany . It is used for N-body/Smoothed-particle hydrodynamics (SPH) simulations in cosmology. The code is released under the GNU GPL license and is suitable for cosmological simulations of structure formation as well as interacting galaxies. It employs the traditional N-body method to evolve self-gravitating collisional fluids and simulates compressible gases using smoothed particle hydrodynamics. Gravity calculations are performed using a tree algorithm and support periodic boundary conditions. The code employs individual and adaptive time steps for all particles, combined with a dynamic tree update scheme. Due to the Lagrangian nature of Gadget, it can span a very large dynamic range in space and time. }
The cosmological parameters employed here are: $\Omega_m=0.308$, 
$\Omega_\Lambda=0.692$,
$\Omega_{b}=0.049$,
$H_0=67.8\,{\rm km}\,s^{-1}\,\rm{Mpc}^{-1}$. 
The dark matter sample contains totally $1024^3$ particles.
The halo sample is generated by the on-the-fly Friends-of-Friends halo finder.
A total of 2,413,300 dark matter halos are found with mass ranging from $3.40 \times 10^{11} M_\odot/h$ to $7.30 \times 10^{14} M_\odot/h$, located at a redshift of $z=1$.

%% file: dmfull.tex
\begin{figure*}[htbp]
\centerline{\includegraphics[width=\fsize]{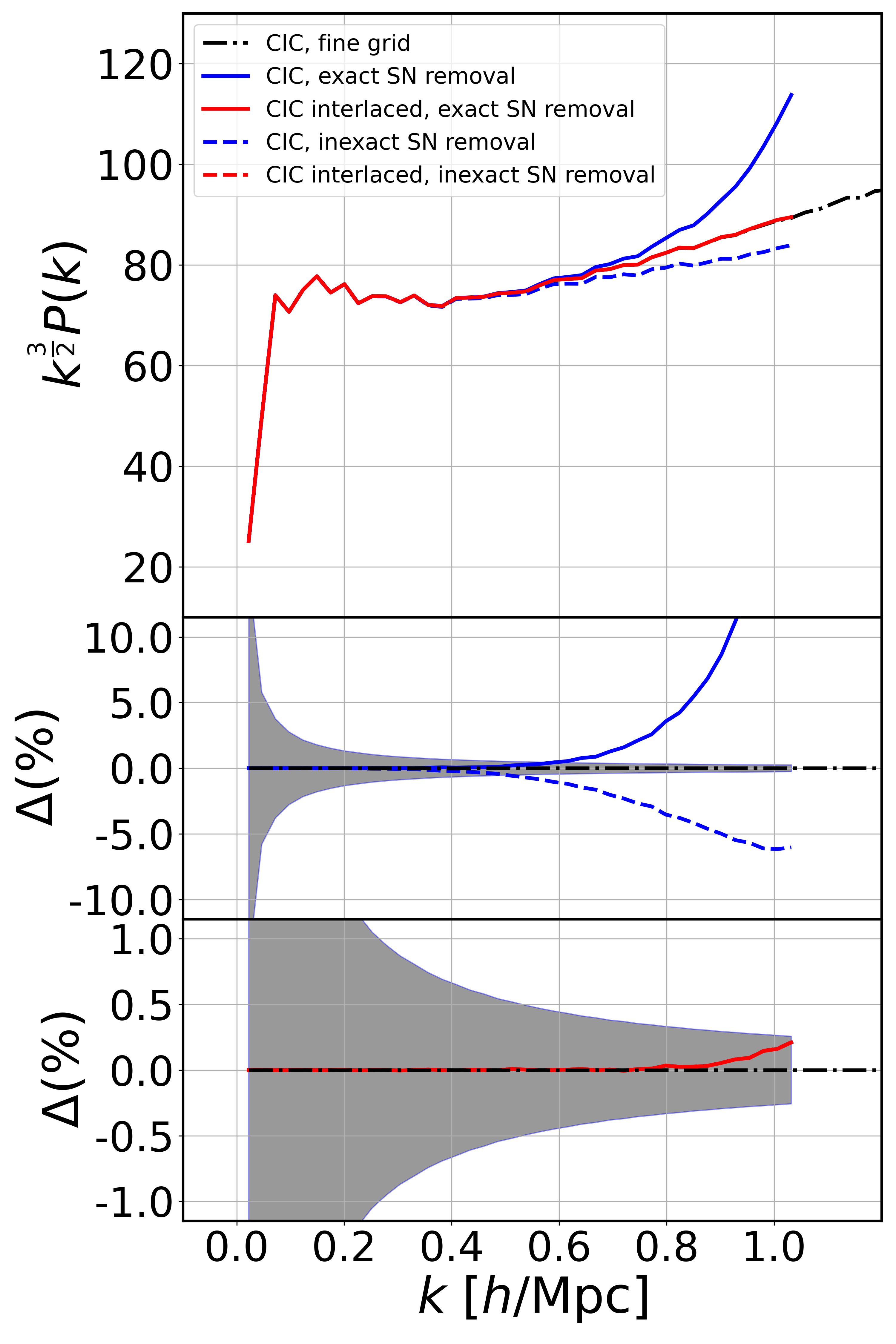}
\includegraphics[width=\fsize]{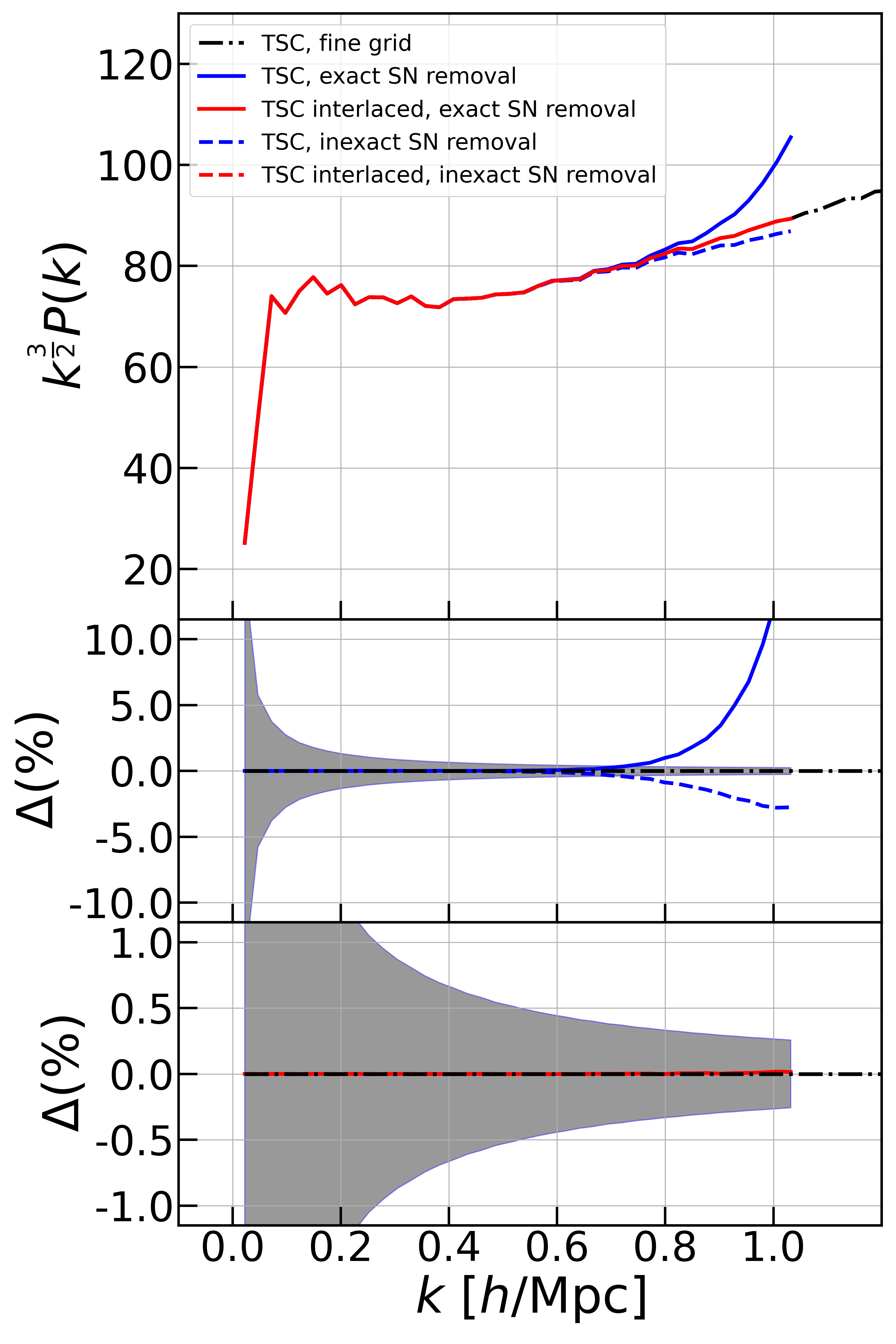}}
\caption{The power spectra of the full dark matter sample, with results for the CIC window function on the left and the TSC window function on the right.
All results extend up to the Nyquist frequency of its own. The top panel shows the power spectrum functions, while the middle and bottom panels depict the ratio of the power spectrum estimates to the results from the high-resolution grid. The blue lines represent the comparison between the two estimation methods without interlacing technique, and the red lines represent the comparison with interlacing technique. The dashed lines are calculated using the \texttt{Nbodykit} library, following the sequence of operations that involve correcting the window function first and then correcting the shot noise, according to the estimation formulae Eq.\,(\ref{eqn:noint-win-shot}) and Eq.\,(\ref{eqn:int-win-shot}). Conversely, the solid lines follow Eq.\,(\ref{eqn:noint-shot-win}) and Eq.\,(\ref{eqn:int-shot-win}). The gray region represents the statistical fluctuations of the power spectrum, i.e., $\pm \sigma=\pm \frac{P+1/n_s}{\sqrt{N_{\textrm{mode}}}}$, where $N_{\textrm{mode}}$ denotes the number of modes of each $k$ bin.}
\label{fig:full_dm}
\end{figure*}
The dark matter particle distribution in simulation is considered to be a Poissonian sampling of the underlying matter field.
Thus we first test the results on the simulated dark matter samples
using the CIC and TSC window functions with a grid size of $N_g = 200$. 
The mesh size is $3\,\mpch$ and the corresponding Nyquist frequency $k_N=1.047\,\hmpc$.
We choose this relatively large mesh size under the consideration that for Stage-IV surveys a cubic box covering the whole footprint may reach $(6\,\gpch)^3$ and the analysis will be performed on $2000^3$ mesh.
We compare the differences among the four estimators mentioned above. 
To reflect the magnitude of systematic biases for the samples, we treat the results from a 10 times higher sampling rate, $N_g = 2000$, as the true values for the power spectrum. 
We have confirmed that, at this high sampling frequency, the power spectrum results from different estimators with different window functions are overlapping with each other below the Nyquist frequency of $N_g=200$ mesh. 
Therefore, we can consider it as the true value for a low sampling frequency result.

For the full sample with a total count of $N_s = 1,073,741,824$ and a number density of $n_s = 4.97\,\hmpct$ , the power spectrum results are shown in Figure~\ref{fig:full_dm}.
Under the CIC window, Eq.\,(\ref{eqn:noint-shot-win}) yields an overestimated result, with an error exceeding 10\% at $k = 0.9k_N$. Eq.\,(\ref{eqn:noint-win-shot}) gives an underestimated result, with an error of around 6\% near $k = k_N$. 
The statistical uncertainty of the band power is given by $\sigma= \frac{P+1/n_s}{\sqrt{N_{\textrm{mode}}}}$, where $N_{\textrm{mode}}$ denotes the number of modes in each $k$ bin.
Note that although we are attempting to clearly subtract the mean shot noise from the measurement, the shot noise residual still contribute to the statistical uncertainty.
Both errors far exceed the level of statistical uncertainty level denoted by grey band. 
Note that the difference between the two estimators fully depends on the different corrections applied to the signal term, since the shot noise level under such circumstances, $\frac{1}{n_s}=0.20\ \mpcht$, is insignificant compared to the signal. 
Yet with the interlacing technique, both estimators Eq.\,(\ref{eqn:int-win-shot}) and Eq.\,(\ref{eqn:int-shot-win}) remain within the statistical fluctuations at the Nyquist frequency.
The overlapping between the two is expected 
since Eq.\,(\ref{eqn:int-win-shot}) and Eq.\,(\ref{eqn:int-shot-win}) is exactly same when shot noise is fully negligible.
The small deviation ($\sim 0.25\%$ at $k=k_N$) from unity is the result of the aliasing effect on the signal, even the interlacing technique is adopted.
In Appendix~\ref{sec:shotnoise_highorder} we provide high-order interlacing schemes which can further reduce the deviation.

Under the TSC window, the systematic biases are reduced. Eq.\,(\ref{eqn:noint-shot-win}) provides an overestimated result, with an error of approximately 18\% at $k = k_N$. Eq.\,(\ref{eqn:noint-win-shot}) yields an underestimated result, with an error of  3\% near $k = k_N$. Both errors still significantly exceed the level of statistical fluctuations. However, with the interlacing technique, both Eq.\,(\ref{eqn:int-shot-win}) and Eq.\,(\ref{eqn:int-win-shot}) exhibit errors much smaller than the level of statistical fluctuations within the Nyquist frequency, and also far below $1\%$. 
%Specifically, Eq.\,(\ref{eqn:int-shot-win}) has an error within 0.015\% at $k = k_N$, while Eq.\,(\ref{eqn:int-win-shot}) overlaps with it with a relative deviation within 0.00001\%.

%The result is to be expected, given that the discrepancy between the two estimators lies solely in the shot noise term. With an high level of number density, further correction on shot noise term(as we suggest in this paper) would be ineffectual/futile.

%(optional)The idea of employing interlacing to reduce aliasing while rectifying window function by $C^{\int}$ is practicable, yet it is inherently against itself. Since such correction assumes
%$\sum_\bfn|W(\bfk+2k_N\bfn)|^2P(\bfk+2k_N\bfn)\approx  P(\bfk)C(\bfk)$
%, which is to amplify aliasing terms instead of ignoring it. The resulting plot is shown in pink, which as you can see the idea of stress aliasing terms rather than ignore it would bring you no good in estimation of power spectrum while interlacing is employed. 

%% file: dm100pc.tex
\begin{figure*}[htbp]
\centerline{\includegraphics[width=\fsize]{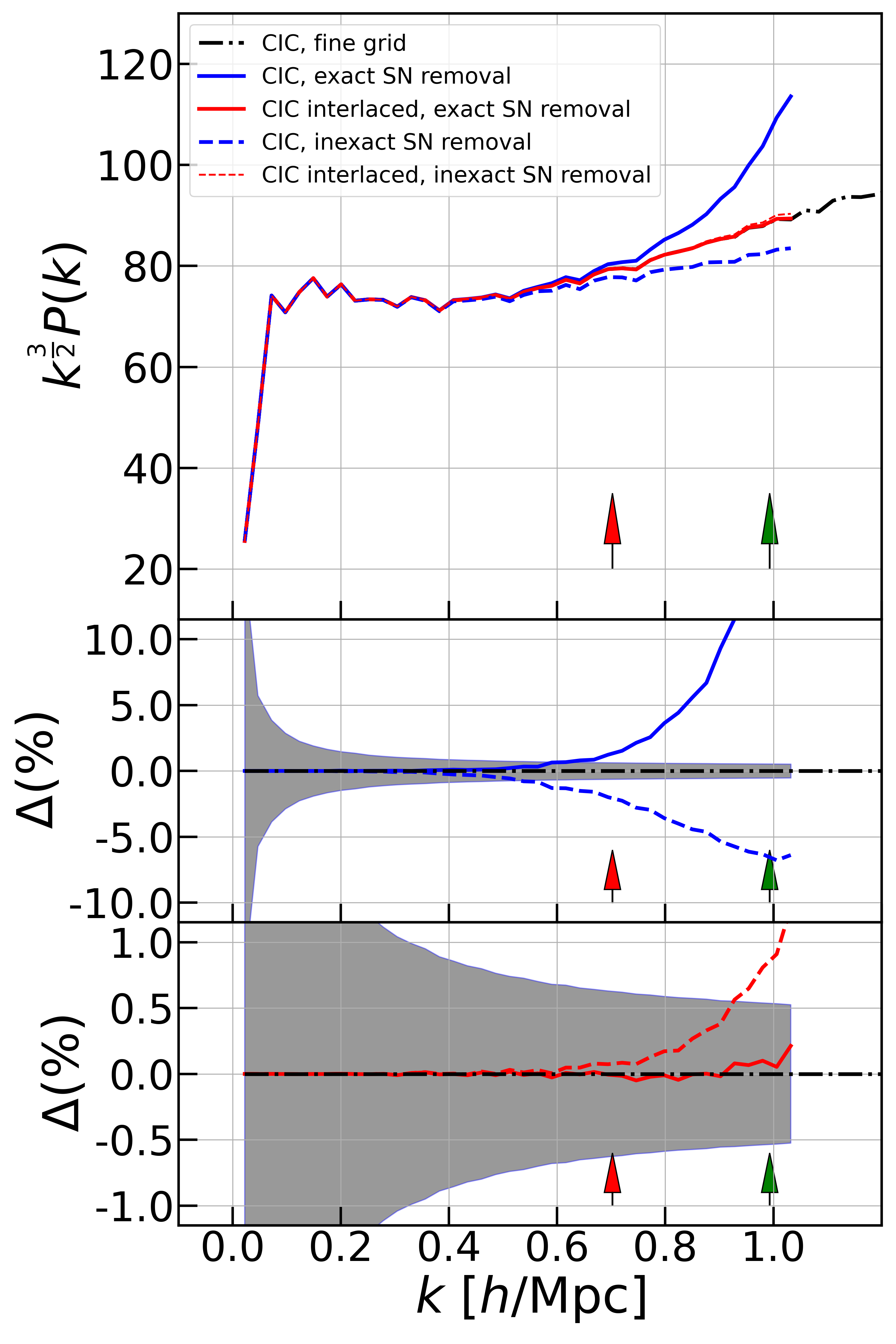}
\includegraphics[width=\fsize]{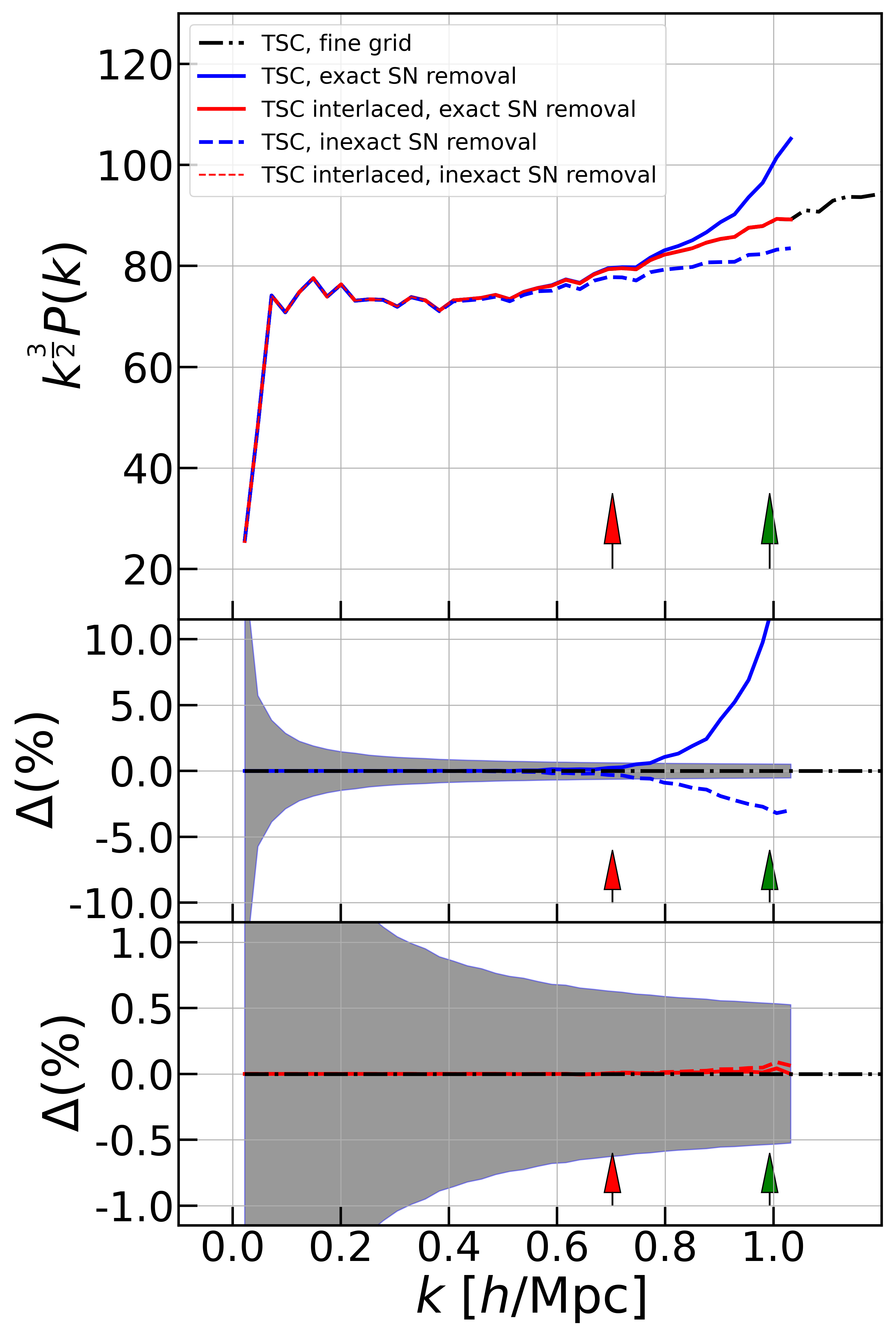}}
\caption{The power spectrum of the dark matter sampled down to $n_s = 1.12\times 10^{-2}\,\hmpct$. The red arrows in the graph indicate the frequency $k = \pi\sqrt[3]{n_s}$ that characterizes the sample's average spatial separation, and the green arrows indicate the scale where shot noise level surpass that of the signals.  Rest of the symbols have the same meaning as those in Figure~\ref{fig:full_dm}.
With interlacing, the estimator Eq.\,(\ref{eqn:int-shot-win}) (red solid line) has much better performance than Eq.\,(\ref{eqn:int-win-shot}) (red dashed line) for CIC window.
}
\label{fig:100percent_dm}
\end{figure*}

To highlight the difference among the various estimators towards low number density limit, we randomly select three subsamples of $N_s = 2,413,300$, $N_s = 241,330$, and $N_s = 24,133$ from the entire dark matter data set.
The sample size can be kind of arbitrary but we specifically chose the above sample sizes for the comparison with later-discussed halo samples of the same sizes.
For the sample with a total count of $N_s = 2,413,300$ and a number density of $n_s = 1.12\times 10^{-2}\,\hmpct$, the power spectrum results are shown in Figure~\ref{fig:100percent_dm}.

For both CIC and TSC window without interlacing, Eq.\,(\ref{eqn:noint-shot-win}) and Eq.\,(\ref{eqn:noint-win-shot}) 
 has similar result as the full sample.
The reason is that the number density here is still sufficiently high, and thus the main problem is the aliasing in the signal term.
When interlacing is adopted, the performance of shot noise subtraction becomes important.
Under the CIC window, Eq.\,(\ref{eqn:int-win-shot}) exhibits errors larger than statistical fluctuations at the scale of $k >0.9 k_N$, reaching around 1\% at $k = k_N$. Meanwhile, Eq.\,(\ref{eqn:int-shot-win}) maintains an error within 0.2\%, and below the statistical uncertainty.
While for TSC window with interlacing, the bias from the two estimators is almost negligible.

%Under the CIC window, Eq.\,(\ref{eqn:noint-shot-win}) yields an overestimated result, with an error exceeding 10\% at $k = k_N$. Eq.\,(\ref{eqn:noint-win-shot}) gives an underestimated result, with an error of around 6\% near $k = k_N$. Both errors far exceed the level of statistical fluctuations. However, with the interlacing technique, Eq.\,(\ref{eqn:int-win-shot}) exhibits errors larger than statistical fluctuations only at the scale of $k = k_N$, reaching around 1\% at $k = k_N$. Meanwhile, Eq.\,(\ref{eqn:int-shot-win}) maintains an error within 0.2\%.

%Under the TSC window, the errors are reduced. Eq.\,(\ref{eqn:noint-shot-win}) provides an overestimated result, with an error of approximately 10\% at $k = k_N$. Eq.\,(\ref{eqn:noint-win-shot}) yields an underestimated result, with an error of around 5\% near $k = k_N$. Both errors still significantly exceed the level of statistical fluctuations. However, with the interlacing technique, both Eq.\,(\ref{eqn:int-shot-win}) and Eq.\,(\ref{eqn:int-win-shot}) exhibit errors smaller than the level of statistical fluctuations within the Nyquist frequency. Specifically, Eq.\,(\ref{eqn:int-shot-win}) has an error within 0.001\% at $k = k_N$, while Eq.\,(\ref{eqn:int-win-shot}) has an error within 0.1\%.

For this number density, we can observe that adopting higher-order window functions with the interlacing technique effectively reduces estimation errors,
which is the main conclusion of \cite{Sefusatti16}.
However, when not using the interlacing technique, the reduction of errors with higher-order window functions is significant but not sufficiently good. 
In terms of performance under the interlacing technique, Eq.\,(\ref{eqn:int-shot-win}) performs better than Eq.\,(\ref{eqn:int-win-shot}).

%% file: dm10pc.tex
\begin{figure*}[htbp]
\centerline{\includegraphics[width=\fsize]{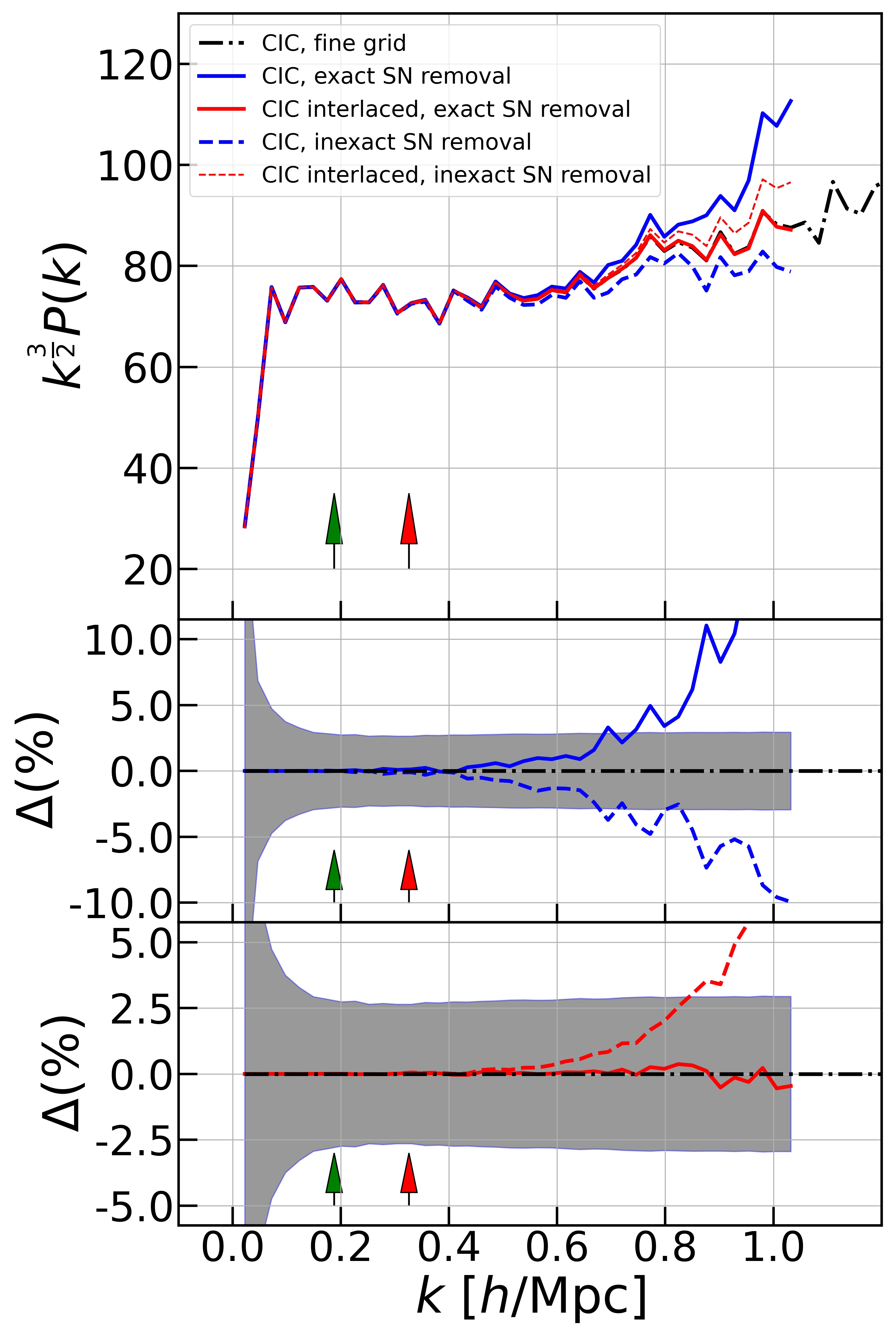}
\includegraphics[width=\fsize]{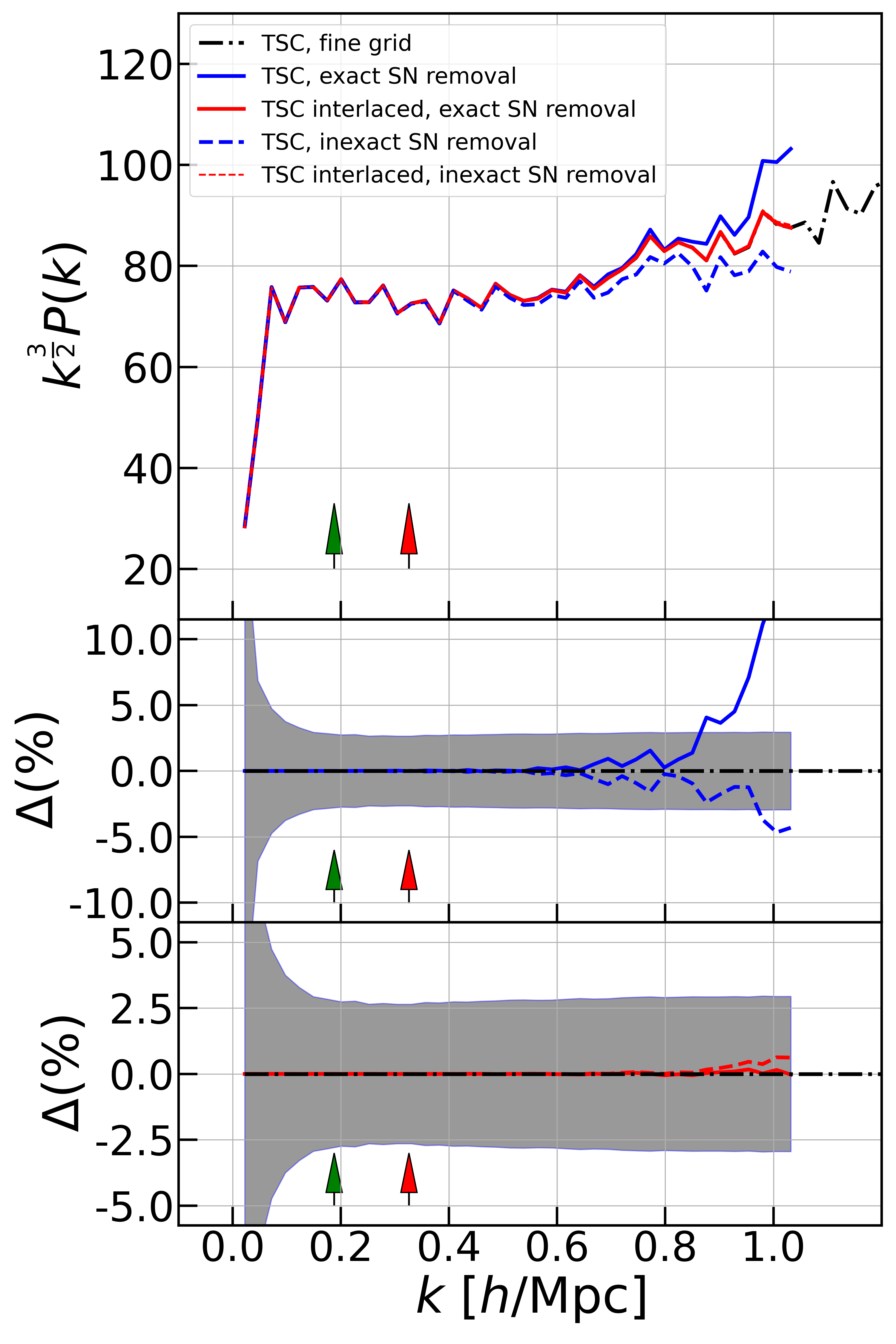}}
\caption{The power spectrum of the dark matter sampled down to $n_s = 1.12\times 10^{-3} \text{ Mpc}^{-3} h^3$. The symbols have the same meaning as those in Figure~\ref{fig:100percent_dm}.
With interlacing, the estimator Eq.\,(\ref{eqn:int-shot-win}) (red solid line) begin to show better result than Eq.\,(\ref{eqn:int-win-shot}) (red dashed line) even for TSC window.
}
\label{fig:10percent_dm}
\end{figure*}

As for the sample with a total count of $N_s = 241,330$ and a number density of $n_s = 1.12\times 10^{-3} \text{ Mpc}^{-3} h^3$, we obtain the power spectrum with the same amplitude as in Figure ~\ref{fig:100percent_dm}, 
but the corresponding shot noise term $\frac{1}{N}$ is ten times larger after down-sampling,
resulting in larger fluctuations on top of the original power spectrum. The results are shown in Figure~\ref{fig:10percent_dm}.

Under the CIC window, Eq.\,(\ref{eqn:noint-shot-win}) provides an overestimated result, with an error reaching 30\% at $k = k_N$. Eq.\,(\ref{eqn:noint-win-shot}) yields an underestimated result, with an error reach 10\% at $k = k_N$. Both errors far exceed the level of statistical fluctuations. 
The increased systematic bias compared to the result in Figure~\ref{fig:100percent_dm} is fully due to the worse shot noise subtraction induced by the estimators, since the underlying aliasing contaminated signal term is the same.
With the interlacing technique, Eq.\,(\ref{eqn:int-win-shot}) exhibits errors exceeding the level of statistical fluctuations around the scale of $k = 0.8k_N$, reaching 10\% at $k = k_N$, while Eq.\,(\ref{eqn:int-shot-win}) maintains an error within 0.6\% despite some fluctuations.
Again, this also explicitly means that Eq.\,(\ref{eqn:int-win-shot}) is far from optimal in terms of shot noise subtraction.

Under the TSC window, the errors are improved. Eq.\,(\ref{eqn:noint-shot-win}) provides an overestimated result, with an error of approximately 18\% at $k = k_N$. Eq.\,(\ref{eqn:noint-win-shot}) yields an underestimated result, with an error of around 5\% near $k = k_N$. Both errors still significantly exceed the level of statistical fluctuations. With the interlacing technique, Eq.\,(\ref{eqn:int-shot-win}) exhibits errors smaller than statistical fluctuations within the Nyquist frequency. Specifically, it has an error below 0.2\% at $k = k_N$.
Meanwhile, the result from Eq.\,(\ref{eqn:int-win-shot}) is also below the statistical uncertainty level, but reach 0.6\% at $k = k_N$.
We emphasize that $n_s = 1.12\times 10^{-3} \text{ Mpc}^{-3} h^3$ is the number density that estimator Eq.\,(\ref{eqn:int-shot-win}) show obvious better performance than Eq.\,(\ref{eqn:int-win-shot}) ($0.2\%$ bias versus $0.6\%$) with TSC window and interlacing technique.

%Compared to the result in Figure ~\ref{fig:100percent_dm}, the systematic bias is significantly larger due to the lower sample density, especially for the estimators with inexact shot noise subtraction.
%We can see that under the interlacing technique, the estimation Eq.\,(\ref{eqn:int-shot-win}) still performs better than Eq.\,(\ref{eqn:int-win-shot}).

%% file: dm1pc.tex
\begin{figure*}[htbp]
\centerline{\includegraphics[width=\fsize]{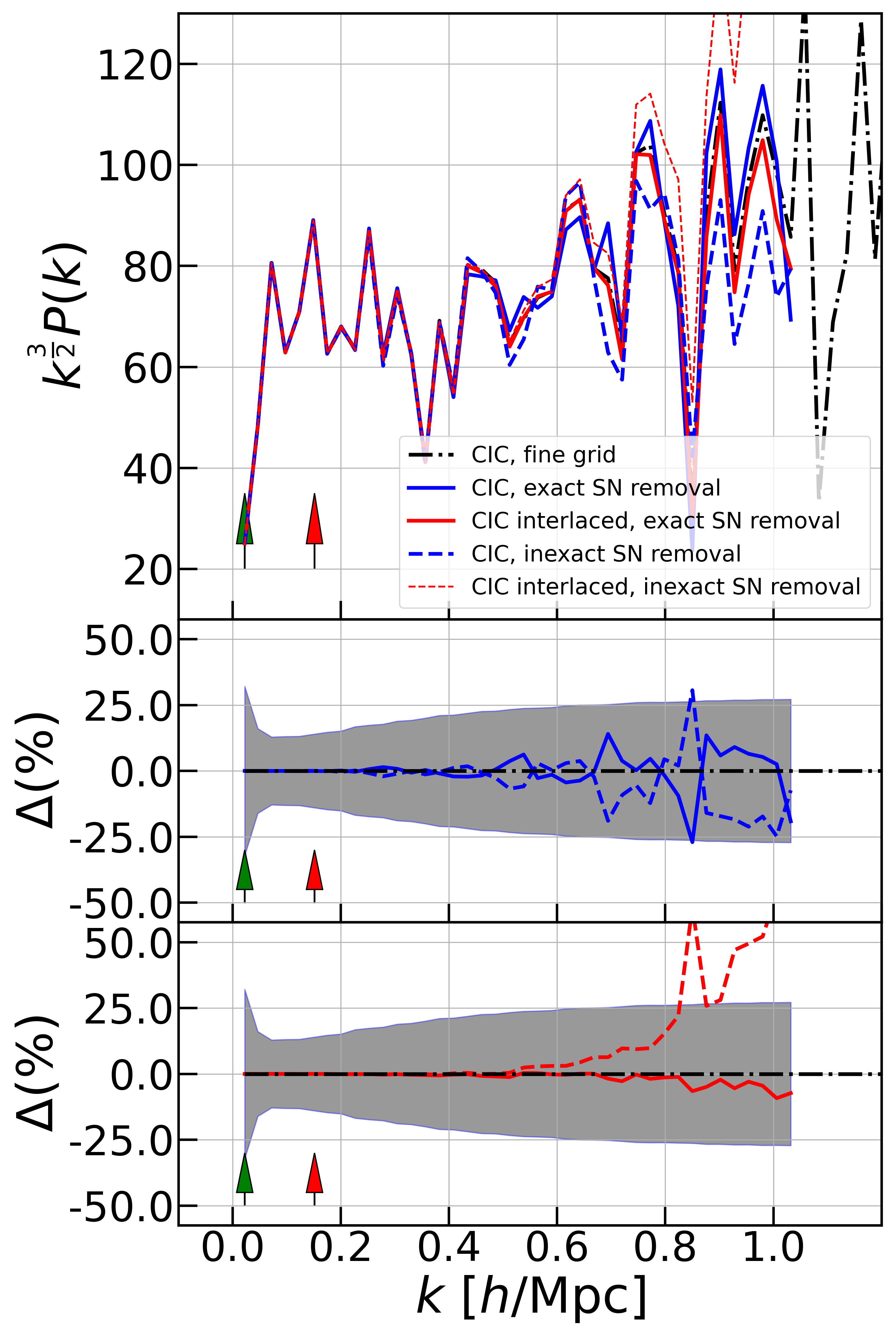}
\includegraphics[width=\fsize]{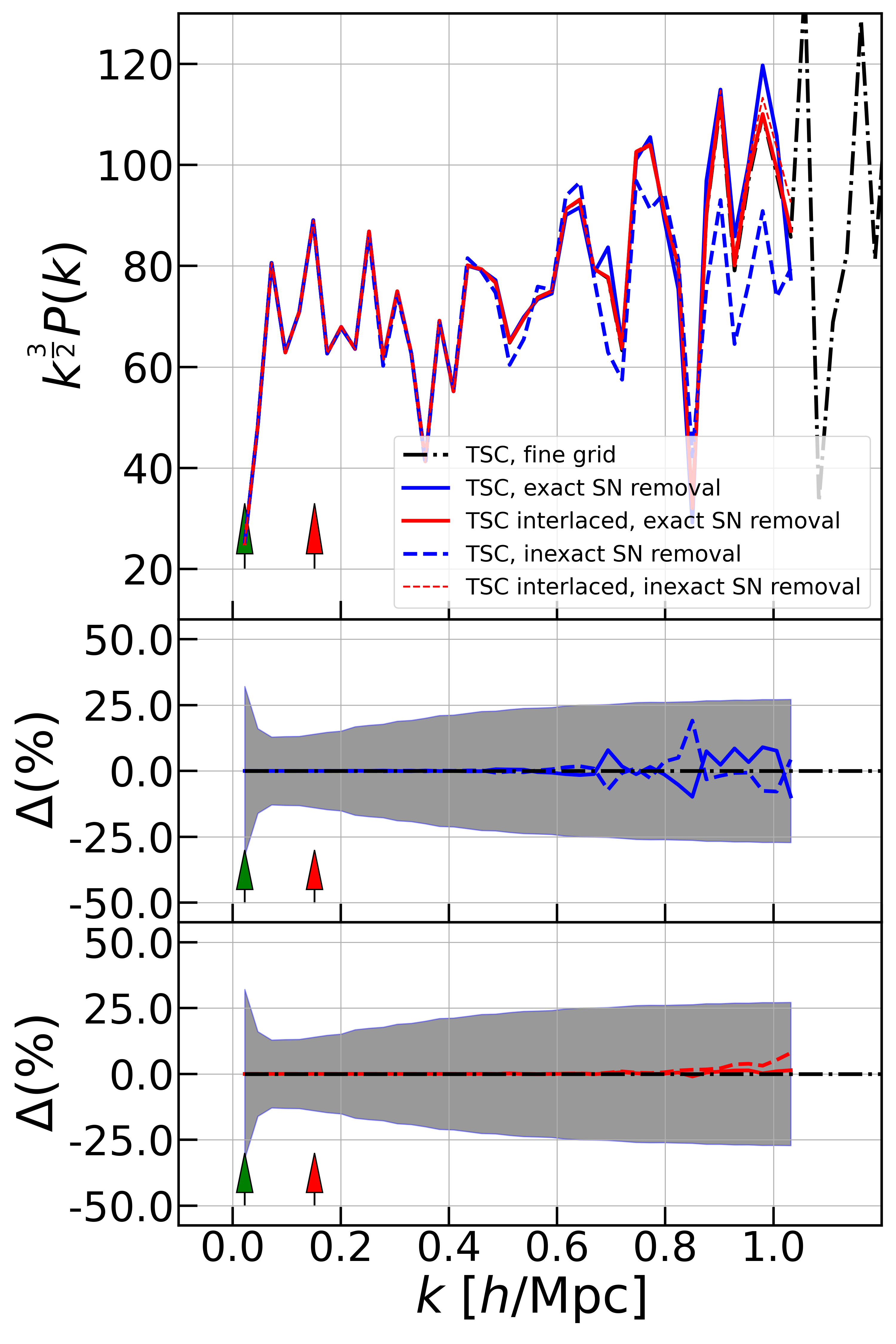}}
\caption{The power spectrum of the dark matter sampled down to $n_s=1.12\times 10^{-4} \,\text{Mpc}^{-3} h^3$. The symbols have the same meaning as those in Figure~\ref{fig:100percent_dm}.}
\label{fig:1percent_dm}
\end{figure*}
As for the sparse sample with a total count of $N_s = 24,133$ and a low number density of $n_s=1.12\times 10^{-4} \,\text{Mpc}^{-3} h^3$, we perform the same computation as described above, and we expect to obtain power spectra with the same amplitude as the former two samples.
However, due to the sparsity of this sample, the resulting power spectrum will exhibit even larger fluctuations compared to the results in Figure~\ref{fig:10percent_dm}. 
The results are presented in Figure~\ref{fig:1percent_dm}.
Under the CIC window, Both Eq.\,(\ref{eqn:noint-win-shot}) and Eq.\,(\ref{eqn:noint-shot-win}) show errors around 10\% for $k>0.65k_N$, which is only half of the statistical fluctuation level.
Under the interlacing technique, Eq.\,(\ref{eqn:int-win-shot}) overestimates the results, exhibits an error of over 50\% at $k=0.9k_N$, while Eq.\,(\ref{eqn:int-shot-win}) underestimates it, exhibits an error below 10\%.
Careful readers may ask why the estimator Eq.\,(\ref{eqn:int-win-shot}) with interlacing technique performs much worse than the estimators Eq.\,(\ref{eqn:noint-win-shot}) and Eq.\,(\ref{eqn:noint-shot-win}) without interlacing.
The reason is that Eq.\,(\ref{eqn:noint-win-shot}) and Eq.\,(\ref{eqn:noint-shot-win}) correctly take into account the aliasing on the shot noise and subtract it (i.e. $\frac{1}{N}C(\bfk)$) from the measurement.  
On the contrary, Eq.\,(\ref{eqn:int-win-shot}) treats the shot noise to be $\frac{1}{N}W^2(\bfk)$ and ignore the aliasing effect on the shot noise.
Thus, the shot noise is greatly underestimated and the residual is large even though the interlacing is adopted.
This example demonstrates that for the shot noise dominated case, the correct subtraction of shot noise is vital.

Under the TSC window, both Eq.\,(\ref{eqn:noint-win-shot}) and Eq.\,(\ref{eqn:noint-shot-win}) give estimations within the size of statistical fluctuations.
However, the fluctuations in the ratio are still obvious.
Under the interlacing technique, again the two estimators have errors smaller than the size of statistical fluctuations, but the above fluctuations in the ratio are less obvious.
Thanks to the large statistical noise level of $25\%$, it seems to be safe to use both estimators.
Note that in terms of the absolute value of the systematic bias,
Eq.\,(\ref{eqn:int-shot-win}) exhibits errors within 2\% near the Nyquist frequency, while Eq.\,(\ref{eqn:int-win-shot}) exhibits errors of around 8\% near the same frequency. 
It can be seen that under this low sampling rate, the estimation using Eq.\,(\ref{eqn:int-shot-win}) still demonstrates a precision advantage.

%% file: halo_mass_order_10percent.tex
\begin{figure*}[htbp]
\centerline{\includegraphics[width=8cm]{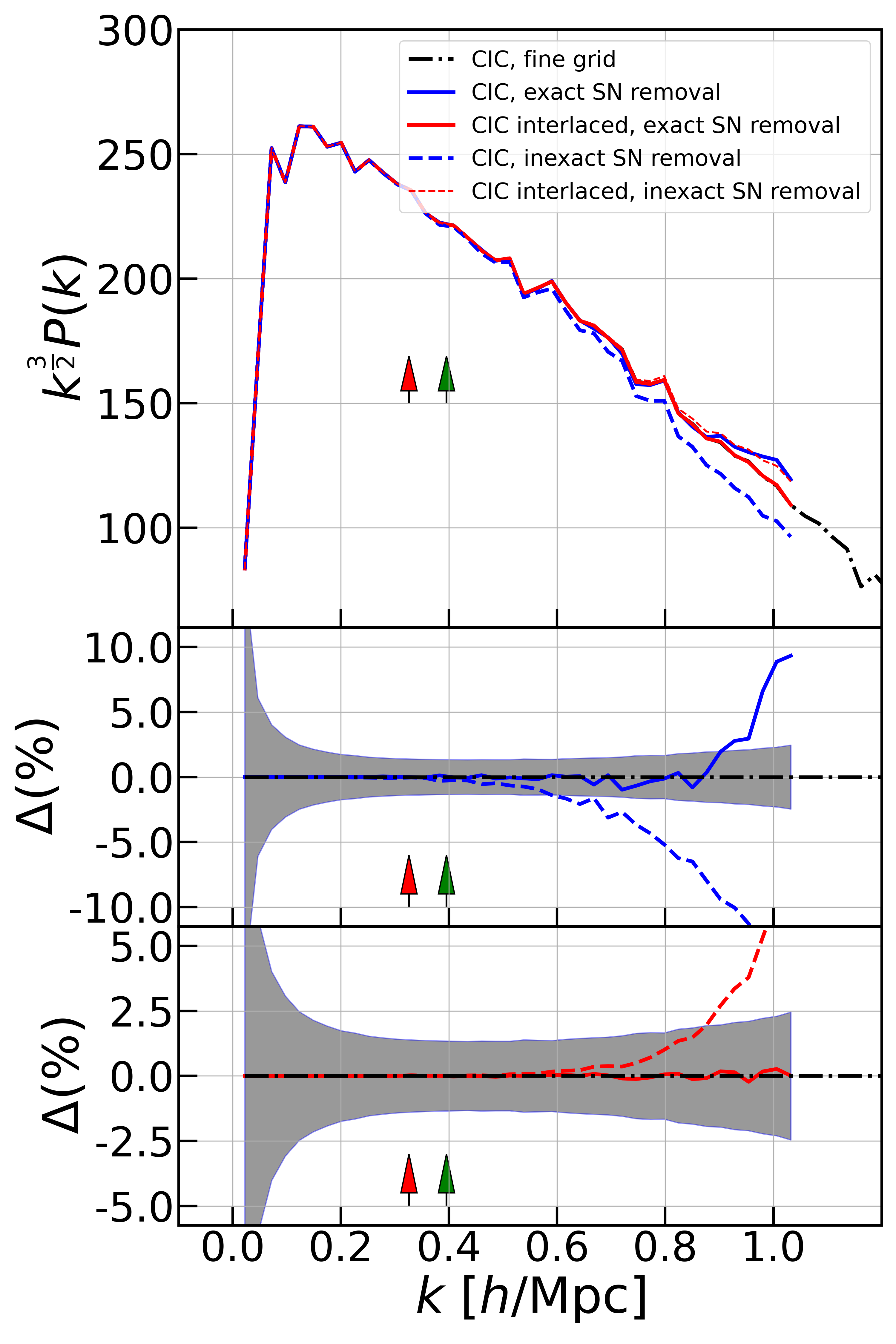}
\includegraphics[width=8cm]{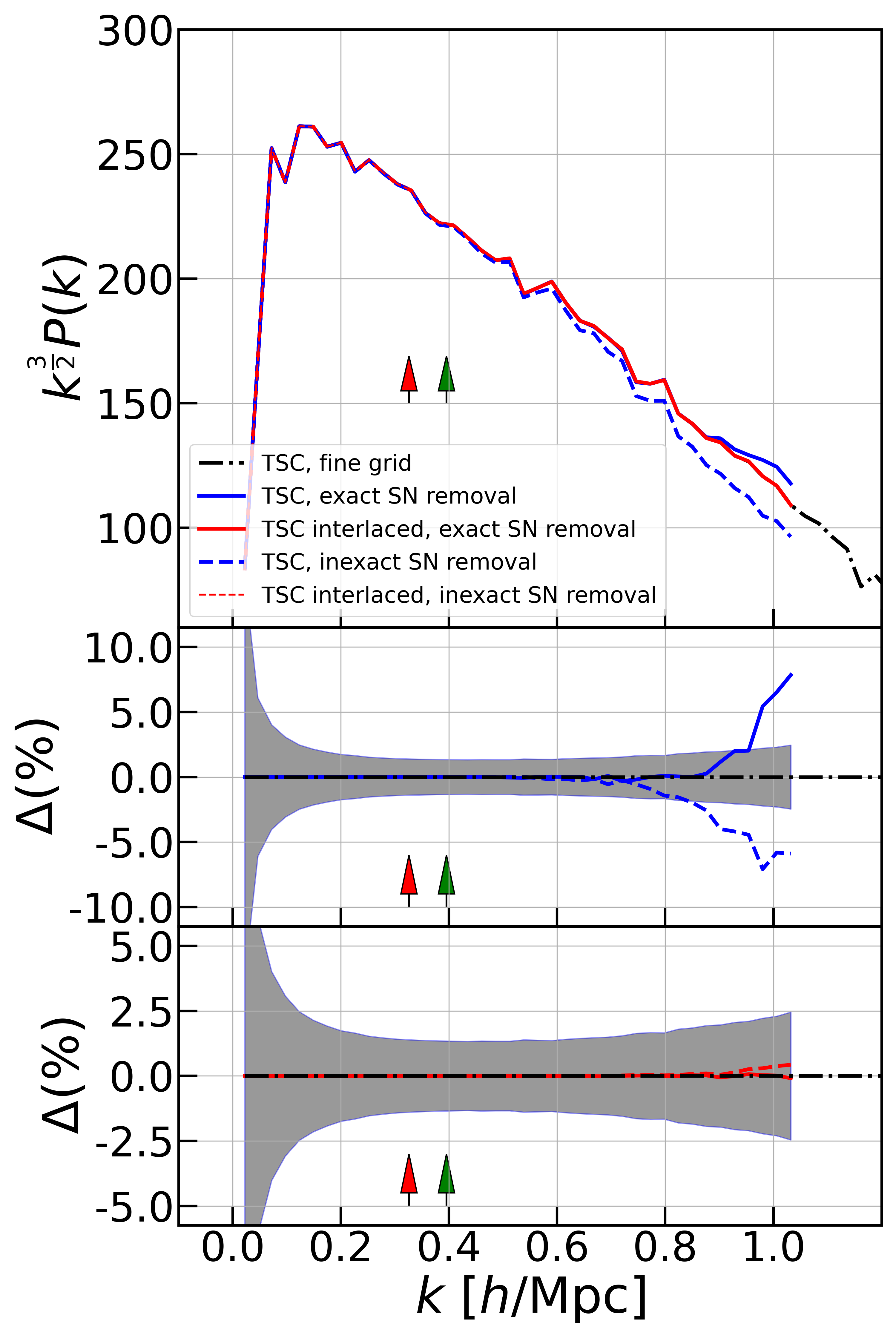}}
\caption{The power spectrum of a high-mass halo subsample ($n_s = 1.12\times 10^{-3} \text{ Mpc}^{-3} h^3$), with the same symbols as in Figure~\ref{fig:100percent_dm}.
For all the estimators, the deviation is slightly smaller than the corresponding dark matter sample, thanks to boosted power amplitude by the halo bias.
}
\label{fig:10percent_inorder}
\end{figure*}

In this section we perform the same calculations as in Section~\ref{sec:dmsample} \& \ref{sec:dmlowsample}, but on the halo samples.
We do not repeat the random down-sampling process on the full halo sample as what we did in Section~\ref{sec:dmlowsample} since the behavior has been well tested there.
Instead, we gradually increase the mass threshold of the haloes to decrease the sample size.
Note that it has been pointed out in the literature (for example, see \cite{Baldauf2013} and \cite{Hamaus10}) that the shot noise contribution in halo power spectrum is in general not Poissonian, particularly for high-mass haloes.
As a consequence, the large scale power is lower than the expected $1/n_s$ and gradually transits to the Poisson expectation in the small-scale.
This means that at least for large scales, the shot noise is over-subtracted in the four estimators we presented.
Here we focus on small scales and this part serves as a performance test on the competition effects between the halo bias which boosts the signal and the shot noise which contaminate the signal.

We first select the top 10\% of the samples in terms of mass from the entire halo sample, resulting in a sample size of $N_s = 241,330$. 
The number density is same as the dark matter sample in Figure~\ref{fig:10percent_dm}, $n_s = 1.12\times 10^{-3} \text{ Mpc}^{-3} h^3$. 
High-mass halos tend to exhibit higher clustering than dark matter.
Therefore, the power spectrum amplitude is boosted.
The large power spectrum amplitude of this sample makes the shot noise less important than the dark matter sample with the same sample size.
%However, the large shot noise term $\frac{1}{N}$ unavoidably amplifies as the average number density decreases, resulting in larger fluctuations on top of the power spectrum. 
The results are shown in Figure~\ref{fig:10percent_inorder}.

Again, Eq.\,(\ref{eqn:noint-shot-win}) provides an overestimated result, and Eq.\,(\ref{eqn:noint-win-shot}) yields an underestimated result for CIC window. Both errors reach approximately 10\% near $k = k_N$, far exceeding the level of statistical fluctuations. 
However, Eq.\,(\ref{eqn:noint-shot-win}) still maintains an error of around 1\% within $k = 0.8k_N$, while Eq.\,(\ref{eqn:noint-win-shot}) exceeds at larger scales. 
Compared to the results in Figure~\ref{fig:10percent_dm}, the systematic deviation is indeed slightly reduced thanks to the halo bias of the sample.
With the interlacing technique, Eq.\,(\ref{eqn:int-win-shot}) exhibits errors larger than statistical fluctuations below the scale of $k = 0.85k_N$, reaching 9\% at $k = k_N$. On the other hand, Eq.\,(\ref{eqn:int-shot-win}) fluctuates but remains within the level of statistical fluctuations. 
Under the TSC window, the errors are improved. Eq.\,(\ref{eqn:noint-shot-win}) provides an overestimated result, with an error of approximately 7\% at $k = k_N$. Eq.\,(\ref{eqn:noint-win-shot}) yields an underestimated result, with an error of around 5\% near $k = k_N$. Both errors still significantly exceed the level of statistical fluctuations. 
%However, with the interlacing technique, Eq.\,(\ref{eqn:int-shot-win}) exhibits errors smaller than statistical fluctuations within the Nyquist frequency. Specifically, it has an error below 0.1\% at $k = k_N$, while Eq.\,(\ref{eqn:int-win-shot}) exceeds statistical fluctuations only near the Nyquist frequency.  We can see that under this sampling scheme, the estimation Eq.\,(\ref{eqn:int-shot-win}) still demonstrates superior accuracy.
With interlacing, both estimators, Eq.\,(\ref{eqn:int-shot-win}) and Eq.\,(\ref{eqn:int-win-shot}), present deviations far less than the statistical uncertainty level.
Careful readers can find that around the Nyquist frequency the estimation Eq.\,(\ref{eqn:int-shot-win}) still demonstrates superior accuracy.

%% file: halo_mass_order_1percent.tex
\begin{figure*}[htbp]
\centerline{\includegraphics[width=\fsize]{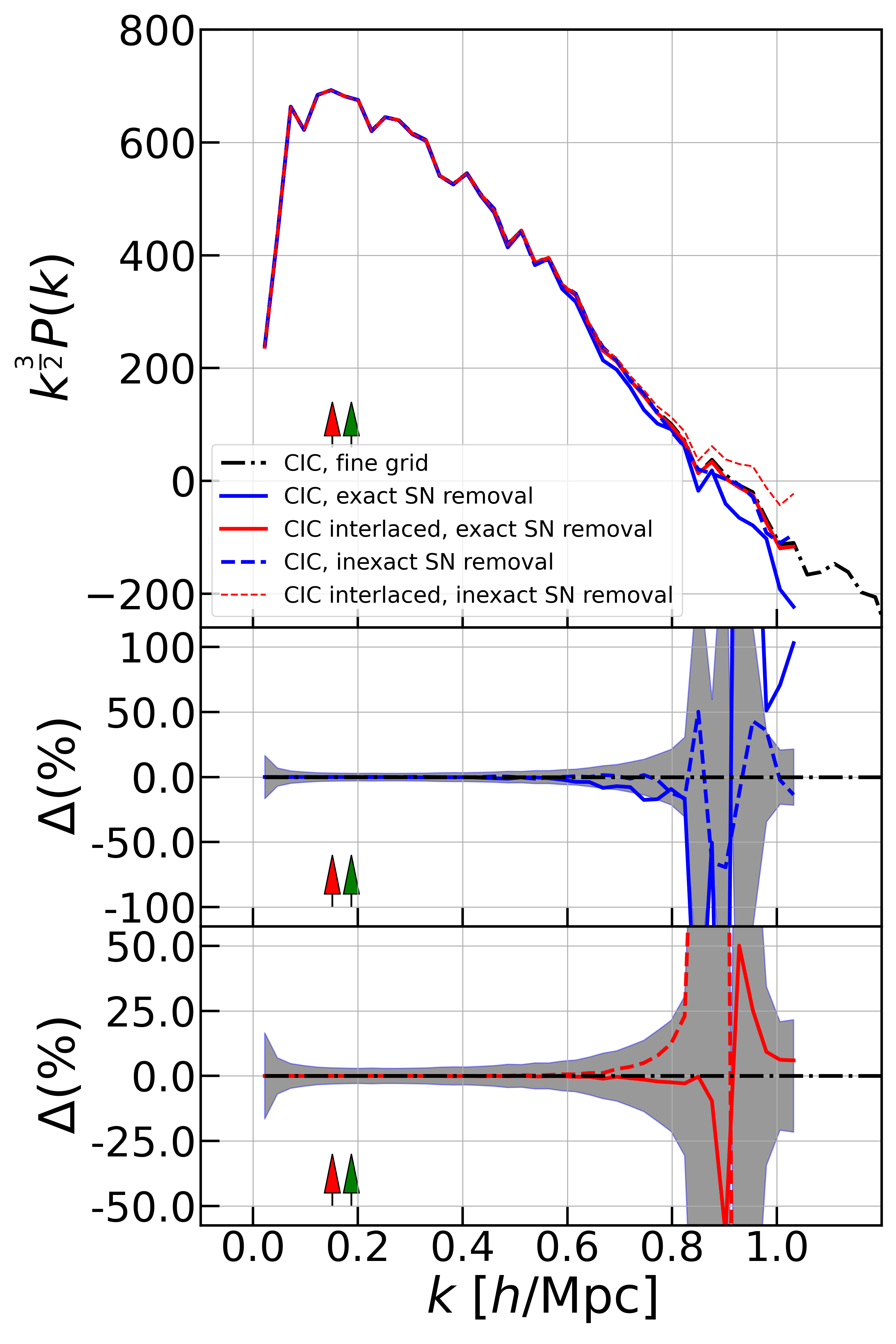}
\includegraphics[width=\fsize]{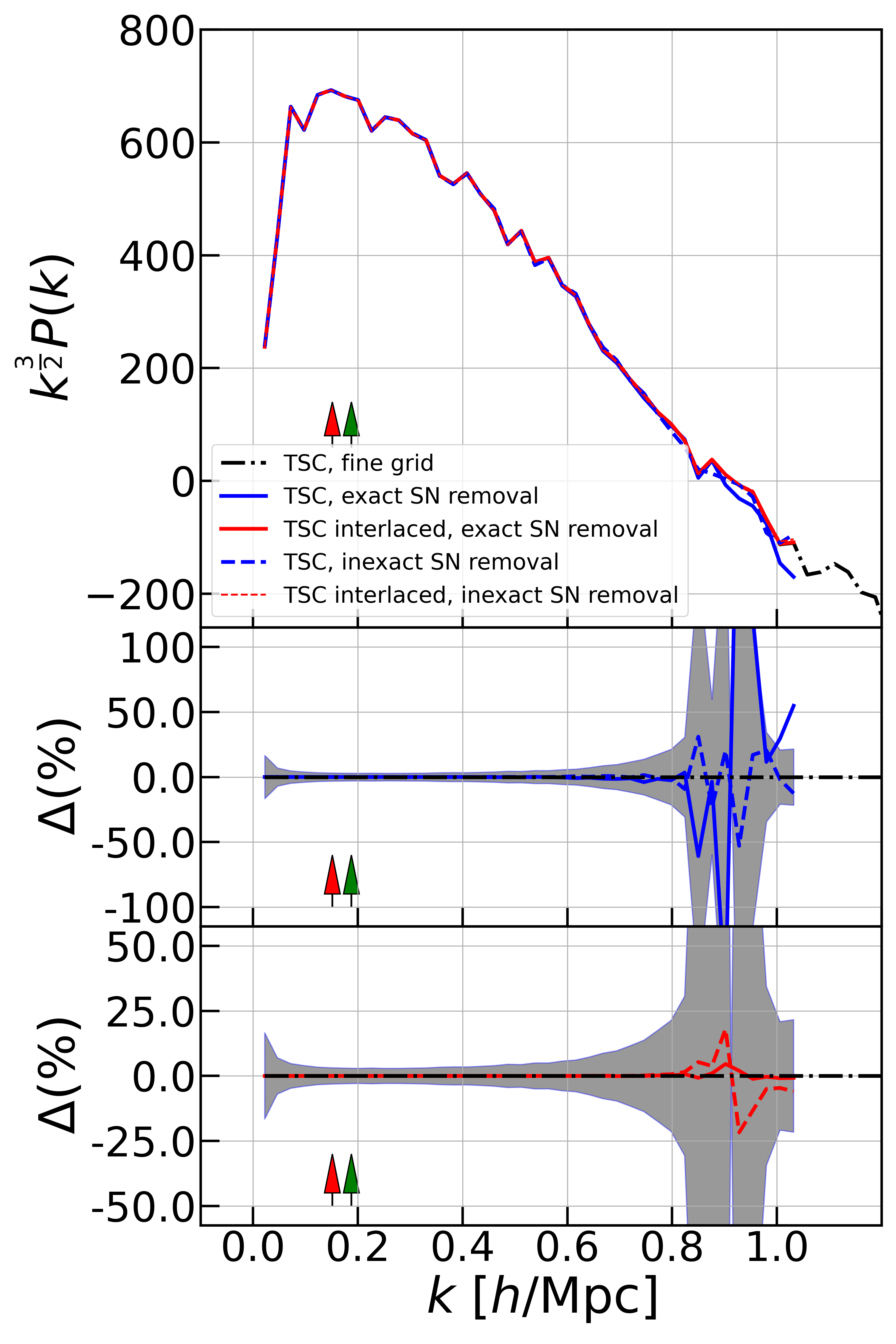}}
\caption{The power spectrum of a $n_s=1.12\times 10^{-4}\, \text{Mpc}^{-3} h^3$ subsample selected in high halo mass, with the same symbols as in Figure~\ref{fig:100percent_dm}.}
\label{fig:1percent_inorder}
\end{figure*}

The power spectrum result for the top 1\% massive halos, with a total sample size of $N_s=24,133$ and a number density of $n_s=1.12\times 10^{-4}\, \text{Mpc}^{-3} h^3$, is presented in Figure~\ref{fig:1percent_inorder}.
The power spectrum exhibits a higher level of clustering due to the selection of high-mass components.
However, the estimated power spectrum is less than zero in the small scale even the fine grid is adopted.
This counterintuitive result is due to the fact that the halo distribution for this sample is sub-Poissonian.
For example, in \cite{Hamaus10}, for the most massive halo bin, the shot noise is only at a level of $2/3$ of the expected $1/N$ value.
Thus, our estimators naively over-subtract the shot noise term by assuming the shot noise being $1/N$.
Despite the negative power in the calculation, the comparison of different estimators is still valid in the sense that they are compared to the fine grid result.
But the zero-crossing in the fiducial power spectrum leads to large fluctuations in the ratio plot.
As a result, the values near that point are sensitive to how the frequency approaches it, leading to a false oscillation in the power spectrum ratio.
We prefer not to compare the values near the zero-crossing scale.
Overall, in the case of the 1\% high mass data, the power spectrum estimation using the interlacing technique with Eq.\,(\ref{eqn:int-shot-win}) still provides the best results.

%% file: conclusion.tex
In this article, we have presented four approaches for power spectrum calculation using FFT and compared their performance. 
It is widely known by the community that the shot noise is a constant $\frac{1}{N}$ as halo/galaxy distribution is Poissonian.
Although this approximation is widely adopted in general estimations (as Eq.\,(\ref{eqn:noint-win-shot}) and Eq.\,(\ref{eqn:int-win-shot})), when using the interlacing technique, the aliasing effects on the signal term are reduced, and thus the impact of aliasing on the shot noise term can no longer be neglected.
We have provided the specific form of the shot noise term under the interlacing technique in Eq.\,(\ref{eqn:interlacedshotnoise_brief}), which demonstrates how the shot noise term is affected by the window function and aliasing effects. 

In precise cosmological analysis, the above accurate shot noise form allows for a more accurate noise subtraction from the measurement, helping to reduce the systematic bias induced by the detailed numerical calculation.
We have pointed out that Eq.\,(\ref{eqn:int-shot-win}) provides a better estimation compared to Eq.\,(\ref{eqn:int-win-shot}). 
For the three dark matter samples with number density ranging from $10^{-2} \text{ Mpc}^{-3} h^3$ to $10^{-4} \text{ Mpc}^{-3} h^3$ analyzed on a coarse mesh of size $3\,\mpch$ with CIC window, the systematic bias from Eq.\,(\ref{eqn:int-shot-win}) is well under control up to the Nyquist frequency.
On the contrary, the bias induced by the estimator Eq.\,(\ref{eqn:int-win-shot}) that ignore the aliasing on the shot noise typically exceeds the statistical uncertainty on the frequency beyond 0.85 times the Nyquist frequency. 
For the cases adopting TSC window, the induced bias from the two estimators is well below the statistical uncertainty.  But Eq.\,(\ref{eqn:int-shot-win}) always shows lower bias than Eq.\,(\ref{eqn:int-win-shot}).

This reduction of the systematics toward the Nyquist frequency also saves the computational cost, since for a good power spectrum estimator such as Eq.\,(\ref{eqn:int-shot-win}) we could use coarse mesh in the analysis for a given precision requirement on a specified scale.
This is of great importance as the precise cosmological constraints rely on large suite of mock observation of order $\mathcal{O}(10^4)$ to obtain the reasonably good covariance matrix~\cite{Yu_2016}.
%\sout{further constrain power spectrum statistics and differentiate small-scale differences between cosmologies.}
%\red{what matters for differentiating cosmologies is the statistical power, not the systematic bias we investigated here.}
Moreover, as realized in literature, the higher-order window functions yield better accuracy, particularly under the interlacing technique. 
Considering the increased computational cost associated with higher-order window functions, it is not recommended to increase the order of the window function without adopting the interlacing technique, as it does not effectively reduce systematic errors. 
Under similar consideration, using better estimator could save computational cost by only using low order window function.
\par
We have made the code implementing Eq.\,(\ref{eqn:noint-shot-win}) and Eq.\,(\ref{eqn:int-shot-win}) available at \url{https://github.com/ColdThunder/FFTps}.

%% file: shotnoise_expansion.tex
\begin{figure*}[htbp]
\centerline{\includegraphics[width=\fsize]{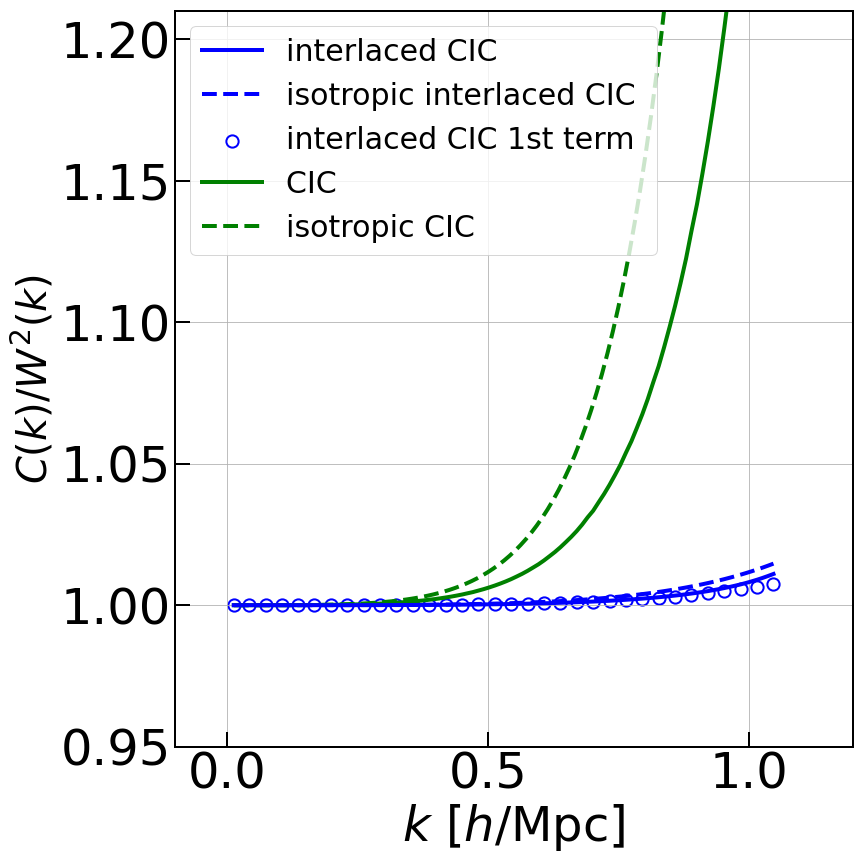}
\includegraphics[width=\fsize]{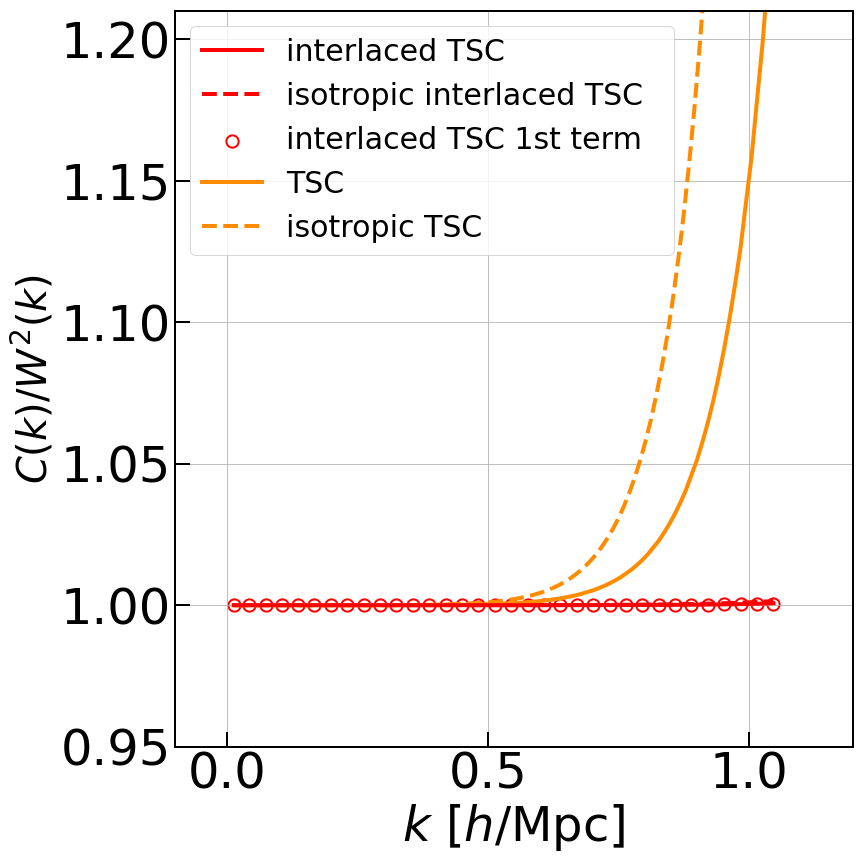}}
\caption{
Analytical shot noise expressions are presented.
On the left side are the results for the CIC window function, and on the right side are the results for the TSC window function.
The dashed lines are the isotropic version following Eq.\,(\ref{eqn:interlacedshotnoise_iso}) for interlacing
and Eq.\,(\ref{eqn:shotnoise}) for non-interlacing.
The solid lines are the exact calculation, following Eq.\,(\ref{eqn:interlacedshotnoise_brief}) and Eq.\,(\ref{eqn:shotnoise_iso}) for interlacing and non-interlacing, respectively.
The data points are the first term in Eq.\,(\ref{eqn:interlacedshotnoise_brief}).}

\label{fig:iso_1stterm}
\end{figure*}

As in \cite{Jing05}, the isotropic version for the shot noise without interlacing is
\begin{equation}
C^{\mathrm{iso}}(k)=\left\{
\begin{aligned}
&1\ , & \text{NGP}\ ,\\
&1-\frac{2}{3}\sin^2\left(\frac{\pi k}{2k_N}\right)\ , & \text{CIC}\ ,\\
&1-\sin^2\left(\frac{\pi k}{2k_N}\right)+\frac{2}{15}\sin^4\left(\frac{\pi k}{2k_N}\right)\ , & \text{TSC}\ ,\\
&1-\frac{4}{3}\sin^2\left(\frac{\pi k}{2k_N}\right)+\frac{2}{5}\sin^4\left(\frac{\pi k}{2k_N}\right)-\frac{4}{315}\sin^6\left(\frac{\pi k}{2k_N}\right)\ , & \text{PSC}\ .\\
\end{aligned}
\right.
\label{eqn:shotnoise_iso}
\end{equation}
In short, it approximates 
$\prod_i S_{2p}[\sin^2(\pi k_i/2k_N)]$ as $S_{2p}[\sin^2(\pi k/2k_N)]$.
For the case with interlacing, the exact anisotropic shot noise is 
Eq.\,(\ref{eqn:interlacedshotnoise_brief}).
A similar approximation can only be straightforwardly applied to the first term $\prod_i C_e(k_i)$.
We show that the first term is overwhelming in Figure~\ref{fig:iso_1stterm}.
Keeping the first term only gives an relative error about 0.3\% when approaching Nyquist frequency with CIC window function, and 0.02\% with TSC window function.
%\sout{To derive shot noise term in 3-dimensional cases, one may extrapolate 1-dimensional cases by simply multiply 3 1-dimensional cases together, turning out to be the first term of Eq.(\ref{eqn:interlacedshotnoise_brief}) which is $\prod_i C_e(k_i)$ only. Such approximation  gives an relative error about 0.3\% while approaching Nyquist frequency with CIC window function, and 0.02\% with TSC window function, as shown in Figure~\ref{fig:first_term}. It is noteworthy that $\prod_i C_e(k_i)$ also constitutes a precise solution within 8 interlaced scheme\red{if you want to say this, we should think of moving the next appendix before this one}, by eliminating odd terms of aliasing in three directions separately, half the number of entire aliasing terms three times simultaneously.}
\par
Given that the above approximation is valid,
it is straightforward to treat the isotropic version of the first term $\prod_i C_e(\pi k_i/4k_N) \approx C_e(\pi k/4k_N)$ as the isotropic approximation for interlacing,
\begin{equation}
C^{\interlace,\mathrm{iso}}(k)=\left\{
\begin{aligned}
&\cos^2\left(\frac{\pi k}{4k_N}\right) , & \text{NGP}\ ,\\
&\cos^4\left(\frac{\pi k}{4k_N}\right)\left[1-\frac{2}{3}\sin^2\left(\frac{\pi k}{4k_N}\right)\right]\ , & \text{CIC}\ ,\\
&\cos^6\left(\frac{\pi k}{4k_N}\right)\left[1-\sin^2\left(\frac{\pi k}{4k_N}\right)+\frac{2}{15}\sin^4\left(\frac{\pi k}{4k_N}\right)\right]\ , & \text{TSC}\ ,\\
&\cos^8\left(\frac{\pi k}{4k_N}\right)\left[1-\frac{4}{3}\sin^2\left(\frac{\pi k}{4k_N}\right)+\frac{2}{5}\sin^4\left(\frac{\pi k}{4k_N}\right)-\frac{4}{315}\sin^6\left(\frac{\pi k}{4k_N}\right)\right]\ , & \text{PSC}\ .\\
\end{aligned}
\right.
\label{eqn:interlacedshotnoise_iso}
\end{equation}
In short, $C^{\interlace,{\mathrm{iso}}}(k)=\cos^{2p}(\pi k/4k_N)C^{\mathrm{iso}}(k/2)$.
Note that the prefactor of $\cos^{2p}x$ can be expressed as polynomial of $\sin^2x$ up to order $p$, leaving the whole formula only containing sine terms.
Or we can express all the terms in the polynomial of $\sin^{2}x$ into polynomial of $\cos^2x$.
However, these further expansions help less when we input the above formula using programming language.
Thus we keep this mixed form of cosine and sine terms in $C^{\interlace,{\mathrm{iso}}}(k)$.
Also in Figure~\ref{fig:iso_1stterm}, the exact shot noise term is compared with the isotropic shot noise for both interlacing and non-interlacing case.
The isotropic approximation for interlacing case is better than the one without interlacing.
%\brown{We show that Under CIC window Eq.\,(\ref{eqn:shotnoise_iso}) gives an overestimation about $0.9$ towards Nyquist frequency, while Eq.\,(\ref{eqn:interlacedshotnoise_iso}) gives an underestimation  about $0.1$. 
%Under TSC window an overestimation about $0.4$ towards Nyquist frequency, while Eq.\,(\ref{eqn:interlacedshotnoise_iso}) gives an underestimation  about $0.07$.} 

%% file: shotnoise_higherorder.tex
\begin{figure*}[htbp]
%\includegraphics[width=8cm]{fig3/chessboard_diagonal_x0.png}\\
%\includegraphics[width=8cm]{fig3/chessboard_diagonal_x1.png}\\
%\includegraphics[width=8cm]{fig3/chessboard_tridim_x0.png}\\
%\includegraphics[width=8cm]{fig3/chessboard_tridim_x1.png}
%\caption{Interlacing scheme in frequency space, while the up two lines represent the diagonal scheme and the last two lines represent the Florian's scheme. The first and the third line shows the scheme's slice when $z=0$ and the second and the fourth line shows the scheme's slice when $z=1$}
\centerline{\includegraphics[width=6cm]{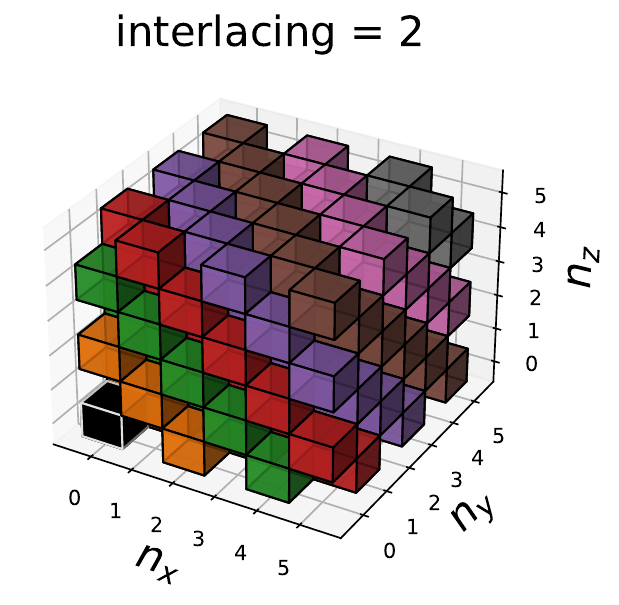}
\includegraphics[width=6cm]{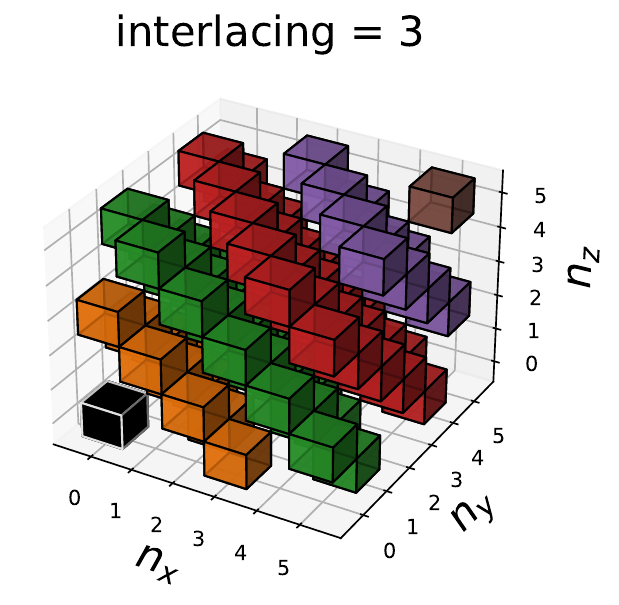}
\includegraphics[width=6cm]{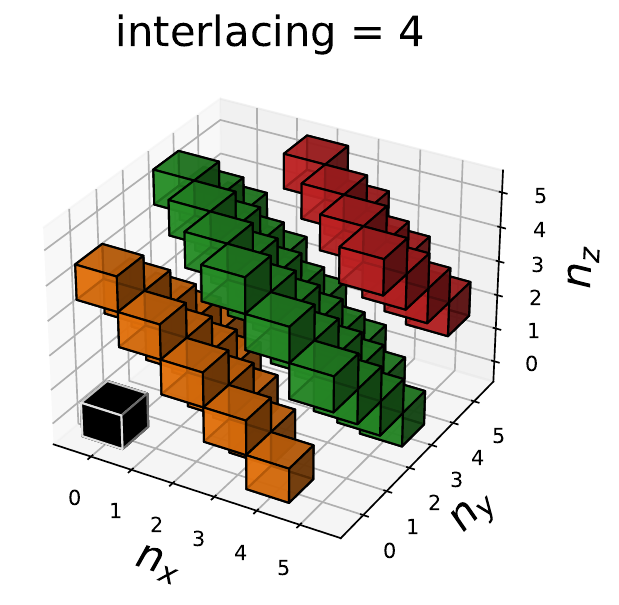}}
\centerline{\includegraphics[width=6cm]{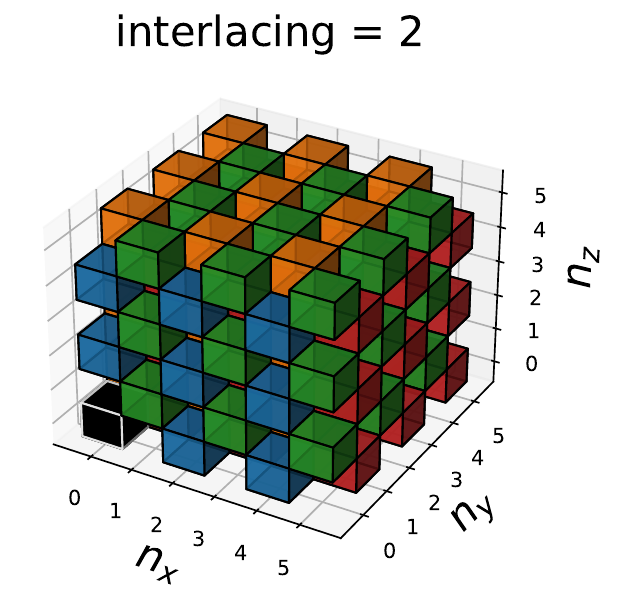}
\includegraphics[width=6cm]{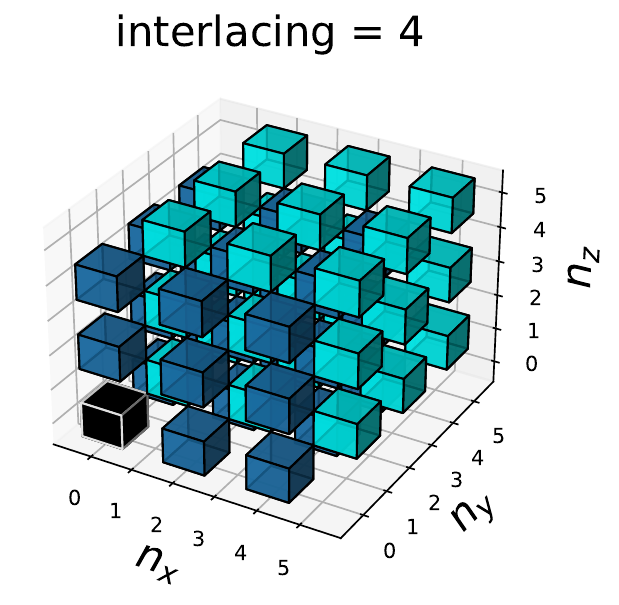}
\includegraphics[width=6cm]{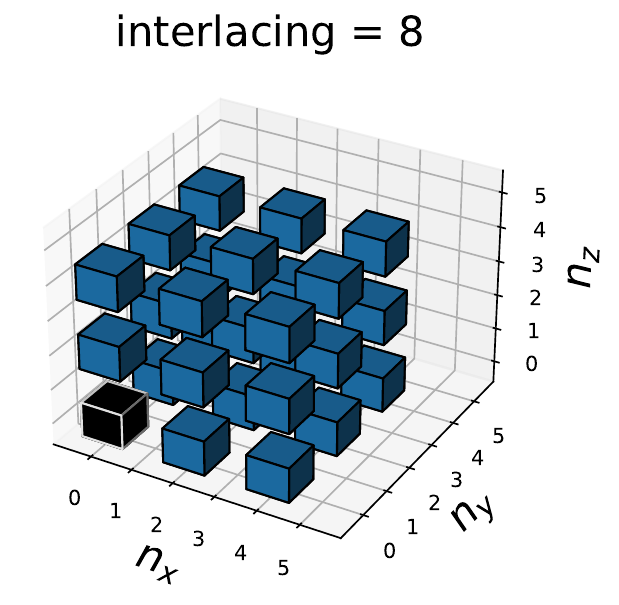}}
\caption{
The interlacing schemes (residual images) in frequency space.
The black voxel denotes the cardinal term with $\bfn=0$. 
The top panels represent the equal-spacing interlacing scheme with 2, 3 and 4 interlaced grids, and the bottom panels represent the bisection interlacing scheme with 2, 4 and 8 interlaced grids. 
In top panels, each color denotes a group of residual images in one plane.
In bottom panels, each color denote a lattice structure that has an offset to the others.
%In top panels, we divided the residual aliased images in the equal-spacing interlacing scheme in different colors as linear-combination terms that's calculable by separating variables, which is the combination of the 8-interlaced schemes with different offsets. 
%Meanwhile the different colors in diagonal schemes marks the periodic structures illustratively, while calculated in the similar way as linearly separating into the combination of the $n^3$-interlaced schemes with different offsets, where $n$ is the order of interlacing. It's noteworthy that both of them possess a same 2-interlaced scheme.
}
\label{fig:Interlacing_scheme}
\end{figure*}

\begin{figure*}[htbp]

\centerline{\includegraphics[width=6cm]{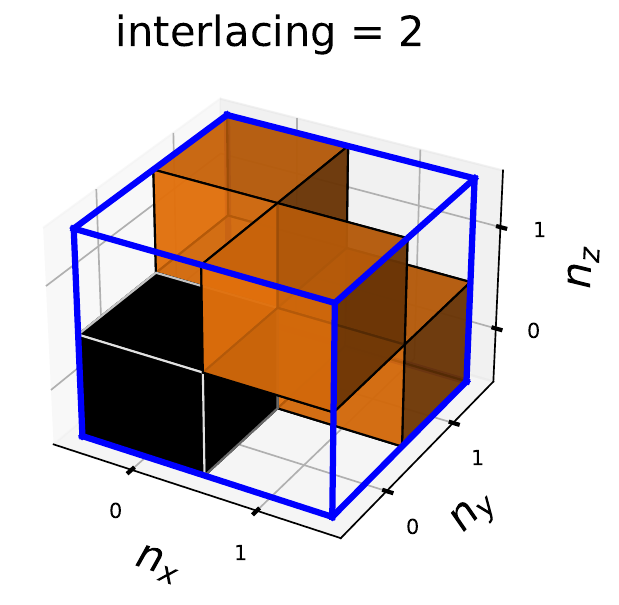}
\includegraphics[width=6cm]{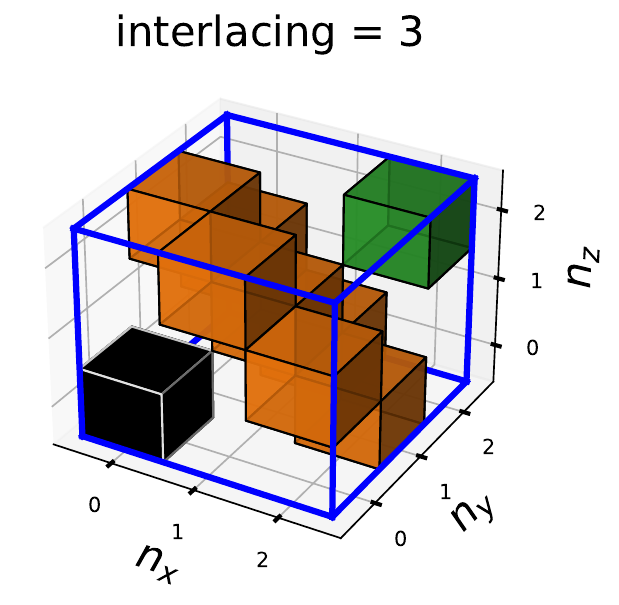}
\includegraphics[width=6cm]{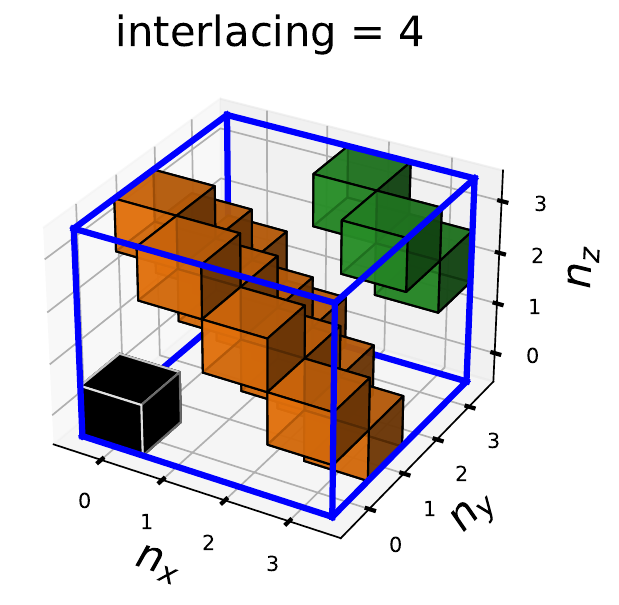}}
\centerline{\includegraphics[width=6cm]{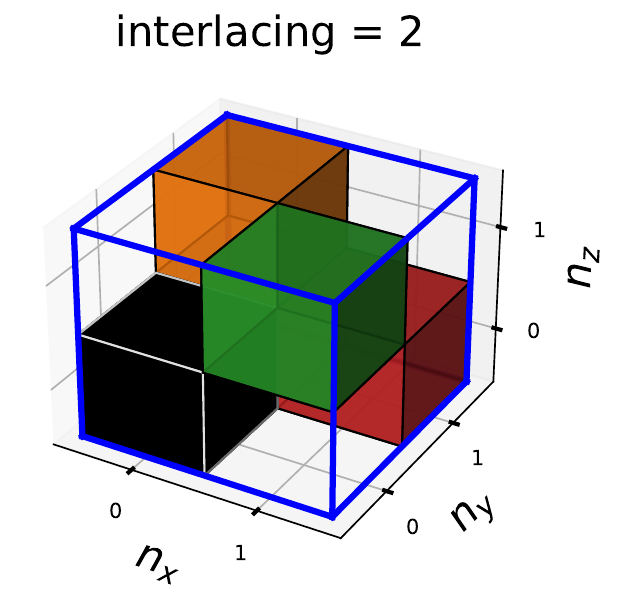}
\includegraphics[width=6cm]{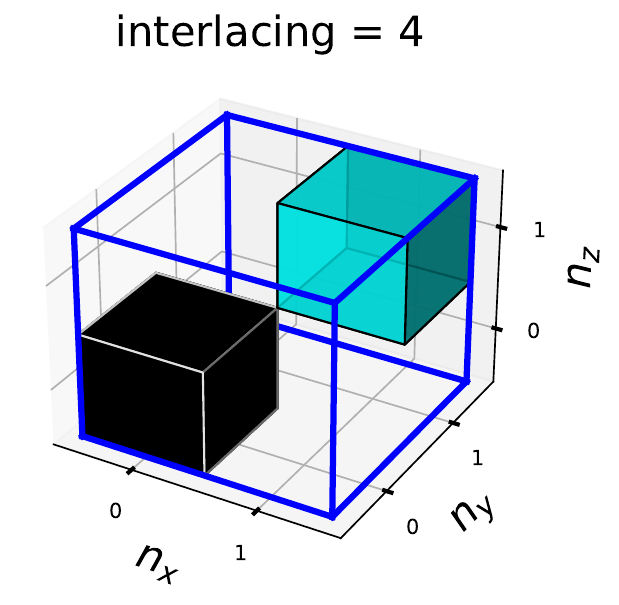}
\includegraphics[width=6cm]{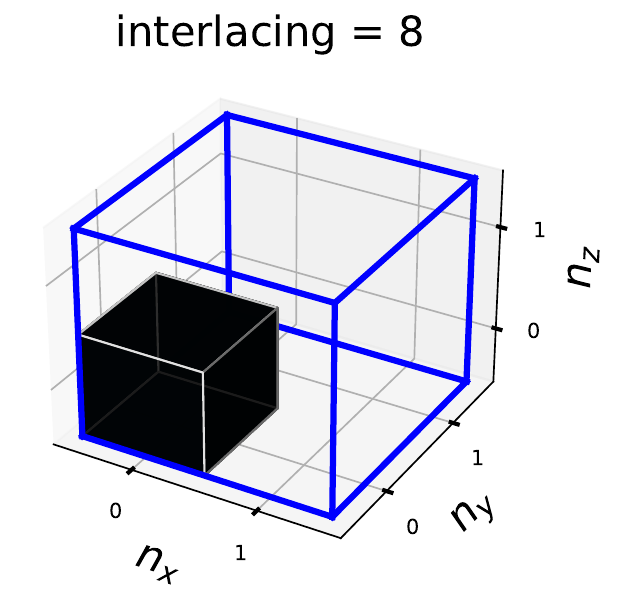}}
\caption{The repeating unit concatenating up in three dimension to make up of the residual images in frequency space in Figure~\ref{fig:Interlacing_scheme}. 
The infinite summation over each block in those units can be calculated individually as $C^{\operatorname{int-1D},n'_x}_m(k_x)C^{\operatorname{int-1D},n'_y}_m(k_y)C^{\operatorname{int-1D},n'_z}_m(k_z)$.
}
\label{fig:Interlacing_scheme_unit}
\end{figure*}

\begin{figure*}[htbp]
\centerline{\includegraphics[width=\fsize]{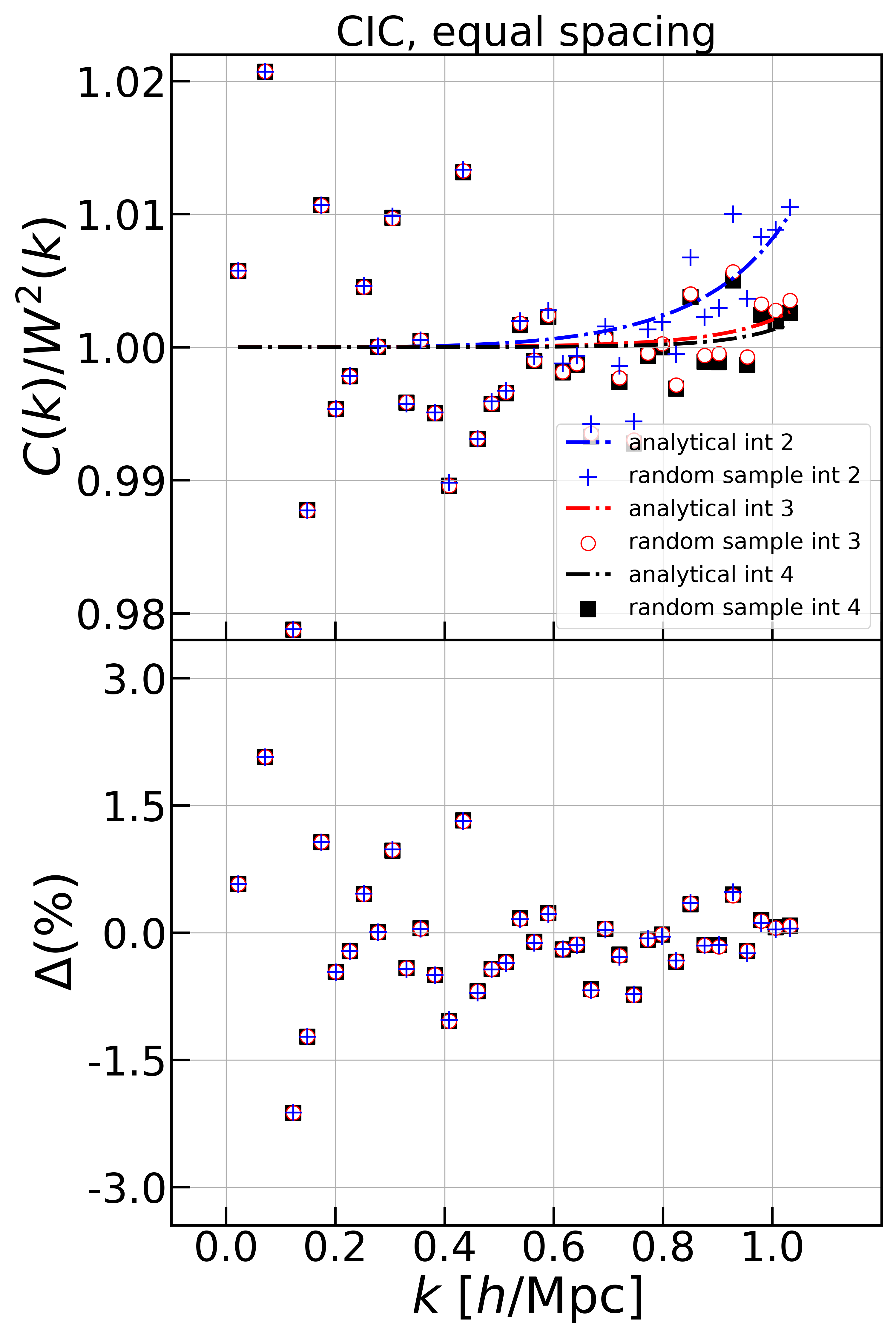}
\includegraphics[width=\fsize]{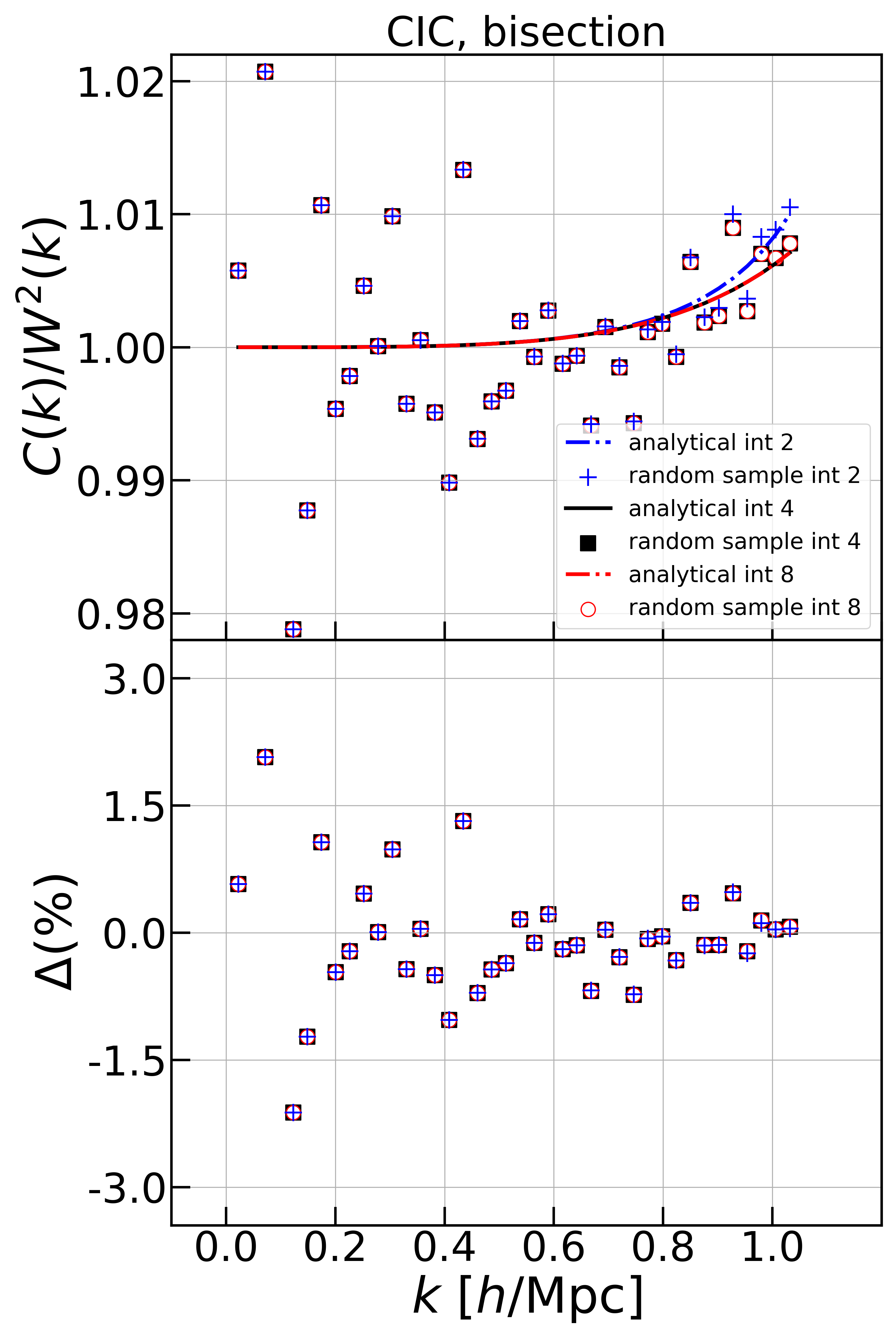}}
\caption{Shot noise estimated from Poisson-distributed random samples with high order interlacing under CIC window function. 
On the left side are the results for the equal-spacing interlacing scheme with 2, 3 and 4 interlaced grids, and on the right side are the results for the bisection interlacing scheme with 2, 4 and 8 interlaced grids.
For the top panels the data points represent the $\langle\delta^f(k)^2\rangle/W^2(k)$ obtained from the random sample, the dot-dashed lines and the solid lines represent the theoretical formulae given by Eq.~(\ref{eqn:interlace_m_shotnoise_pypower}), Eq.~(\ref{eqn:interlace_shotnoise_3m}), Eq.~(\ref{eqn:interlacedshotnoise_3m1}) and Eq.~(\ref{eqn:interlacedshotnoise_3m2}), respectively.
The bottom panels show the deviation between the individual results and their respective theoretical formula. 
The analytical expressions describe the shot noise accurately, leaving only the same sample variance scattering around unity.
}
\label{fig:shot_show_highorder}
\end{figure*}

The main goal of interlacing technique is to eliminate aliased images.
For 2 interlaced grids, the second mesh which has a shift $(H/2, H/2, H/2)$ relative to the primary one can eliminate the main aliasing contamination from $|\bfn|=1$.
To eliminate more images, using more interlaced grids, i.e. high order interlacing, is useful.
However, different high order schemes appear in literature.
We list two of them.
\begin{itemize}
    \item Equal-spacing interlacing scheme.  
    \item Bisection interlacing scheme.
\end{itemize}
\par
For the equal-spacing interlacing scheme,
$m$ interlaced grids are adopted, which are placed at $(jH/m, jH/m, jH/m)$ for $j=0,1,\cdots,m-1$.
This scheme was proposed by \cite{CHEN74} (see also \cite{hockney21}), and used in \texttt{pypower}\footnote{\url{https://github.com/cosmodesi/pypower}, commit 072753e}.
For the bisection interlacing scheme,
the high order interlaced grids are located at the middle of the low order ones.
For 4 interlaced grids, the origin of the interlaced grids are located at
$(0, 0, 0)$,
$(0, H/2, H/2)$,
$(H/2, 0, H/2)$,
$(H/2, H/2, 0)$.
For 8 interlaced grids, the positions are
$(0, 0, 0)$,
$(0, 0, H/2)$,
$(0, H/2, 0)$,
$(0, H/2, H/2)$,
$(H/2, 0, 0)$,
$(H/2, 0, H/2)$,
$(H/2, H/2, 0)$,
$(H/2, H/2, H/2)$.
This scheme appeared in \cite{Florian23} (see also \cite{hockney21}).

The residual aliased images for the two schemes are presented in Figure \ref{fig:Interlacing_scheme}.
Here we directly comment on the two schemes from the sketch in the figure, and leaving the mathematical derivation later.
For the equal-spacing scheme, the elimination seems to be efficient when $m\ge 3$.
We caution that the plot is misleading since we put the $|\bfn|=0$ block in the corner.  
The residual images are distributed anisotropically, and 
all the images on $n_x+n_y+n_z=0$ plane is never eliminated for any order $m$.
Thus, the main contamination is always from eight images at $(\pm 1, \mp 1, 0)$, $(\pm 1, 0, \mp 1 )$ and $(0, \pm 1, \mp 1)$ even for $m\ge 3$.
For bisection interlacing scheme with 4 interlaced grids, the main contamination is from the eight images that share the eight vertices of the primary one.
While moving to 8 interlaced grids, all the images with distance less than 2, i.e. all the neighbour 26 images, are eliminated.

\subsection{Equal-spacing interlacing scheme}
Considering such an interlacing scheme that $m$ girds sequentially shift $(H/m, H/m, H/m)$ to the previous one,
i.e. the $j$th grid ($j=2, 3, \cdots, m$) is shifted from the first one by a distance of $((j-1)H/m, (j-1)H/m, (j-1)H/m)$ and produces a density field $\delta _j^f(\bfk)$.
The arithmetic average of these samplings gives the interlaced denstiy field

\begin{equation}
\begin{aligned}
\tilde\delta^f(\bfk)&=\frac{1}{m}\left[\sumjm \delta^f_j(\bfk)\right]\\
%&=\frac{1}{m}\left[\sumjm \delta^f_j(\bfk)\  e^{-\mathrm{i}j(k_x+k_y+k_z) H/m}\right]\\
%&=\frac{1}{m}\left[\sumjm \delta^f_j(\bfk+2k_N\bfn)\ e^{-2\mathrm{i}\pi j(n_x+n_y+n_z) /m}\right]\\
&=\sum_\bfn{\theta_\bfn{\delta}^f(\bfk+2k_N\bfn)}\ .
\end{aligned}
\label{eqn:interlace_n}
\end{equation}
Similar to Eq.\,(\ref{eqn:interlace_index}),
$\theta_\bfn$ is defined as
\begin{equation}
\begin{aligned}
\theta_\bfn &\equiv \frac{1}{m}\left[\sumjm e^{-2\mathrm{i}\pi j(n_x+n_y+n_z) /m}\right]\\
&=\left\{
\begin{array}{ll}
1\ , & (n_x+n_y+n_z)/m \text{ is an integer}\ ,\\
0\ , & \text{ otherwise}\ .
\end{array}\right.
\end{aligned}
\label{eqn:interlace_m_index}
\end{equation}
Obviously, the residual images locate on planes with $n=n_x+n_y+n_z=ml$ and $l$ being any integer, as in the top panels in Figure \ref{fig:Interlacing_scheme}.

To sum up the contributions from these anisotropically distributed aliased images,
we first decompose the whole space into periodical patches along three Cartesian directions.
Each patch contains same $m\times m$ voxels as shown in the top panels in Figure~\ref{fig:Interlacing_scheme_unit}.
Consequently, the contribution from these patches can be calculated separately for each dimension, 
and inside each patch we need to sum up the contributions from the residual images shown as colored blocks.
Mathematically, this decomposition corresponds to expanding $\theta_\bfn$ as
\begin{equation}
\begin{aligned}
\theta_\bfn &\equiv \frac{1}{m}\left[\sumjm e^{-2\mathrm{i}\pi j(n_x+n_y+n_z) /m}\right]\\
&=\sum_{(n'_x,n'_y,n'_z)\in G}
\left[\frac{1}{m}\sumjm e^{-2\mathrm{i}\pi j(n_x-n'_x) /m}\right]
\left[\frac{1}{m}\sumjm e^{-2\mathrm{i}\pi j(n_y-n'_y) /m}\right]
\left[\frac{1}{m}\sumjm e^{-2\mathrm{i}\pi j(n_z-n'_z) /m}\right]\ .
\end{aligned}
\label{eqn:interlace_m_index_expansion}
\end{equation}
Here, $G$ is the group satisfying $n'_x+n'_y+n'_z=0,m$ for $m=2$, and 
$n'_x+n'_y+n'_z=0, m, 2m$ for $m\ge 3$.
$n'_x$, $n'_y$ and $n'_z$
are integers in between $[0,m)$.
The summation over group $G$ counts the contributions from all residual images in one patch.
As shown in Figure~\ref{fig:Interlacing_scheme_unit}, there are 4 blocks to count for $m=2$, 9 blocks for $m=3$, and 16 blocks for $m=4$.
The summation over $\bfn$ in Eq.~(\ref{eqn:interlace_n}) collects the contributions over all patches.
%As shown in Figure~\ref{fig:Interlacing_scheme_unit}, these residual images are all located on the planes parallel to $n_x+n_y+n_z=0$, spacing with distance $m$ in one dimension.

In one dimension, the shot noise over images locating at $(0, \pm m, \pm 2m, \cdots)$ is
\begin{equation}
\begin{aligned}
C^{\operatorname{int-1D}}_m(k) &= |W(k)|^2\sum_{n=ml} \left(\frac{k}{k-2nk_N}\right)^{2p}
= |W(k)|^2\sum_l \left(\frac{k}{k-2mlk_N}\right)^{2p}\\
&= |W(k)|^2\sum_{l}    
\left(\frac{k/m}{k/m-2lk_N}\right)^{2p}
= \frac{|W(k)|^2}{|W(k/m)|^2}C(k/m)\\
%&=\left\{
%\begin{aligned}
%&\frac{\sin^2\left(\frac{\pi k}{2k_N}\right)}{m^2\sin^2\left(\frac{\pi k}{m2k_N}\right)}\ , &\text{NGP}\ ,\\
%&\frac{\sin^4\left(\frac{\pi k}{2k_N}\right)}{m^4\sin^4\left(\frac{\pi k}{m2k_N}\right)}\left[1-\frac{2}{3}\sin^2\left(\frac{\pi k}{m2k_N}\right)\right]\ , &\text{CIC}\ ,\\
%&\frac{\sin^6\left(\frac{\pi k}{2k_N}\right)}{m^6\sin^6\left(\frac{\pi k}{m2k_N}\right)}\left[1-\sin^2\left(\frac{\pi k}{m2k_N}\right)+\frac{2}{15}\sin^4\left(\frac{\pi k}{m2k_N}\right)\right]\ , &\text{TSC}\ ,\\
%&\frac{\sin^8\left(\frac{\pi k}{2k_N}\right)}{m^8\sin^8\left(\frac{\pi k}{m2k_N}\right)}\left[1-\frac{4}{3}\sin^2\left(\frac{\pi k}{m2k_N}\right)+\frac{2}{5}\sin^4\left(\frac{\pi k}{m2k_N}\right)-\frac{4}{315}\sin^6\left(\frac{\pi k}{m2k_N}\right)\right]\ , &\text{PSC}\ .\\
%\end{aligned}
%\right.\\
&=\frac{\sin^{2p}\left(\frac{\pi k}{2k_N}\right)}{m^{2p}\sin^{2p}\left(\frac{\pi k}{2mk_N}\right)}S_{2p}\left[\sin^2\left(\frac{\pi k}{2mk_N}\right)\right]\ .
\end{aligned}
\label{eqn:interlace_m_shotnoise}
\end{equation}
Straightforwardly, the summation over images locating at $(n_{\textrm{shift}}, n_{\textrm{shift}}\pm m, n_{\textrm{shift}}\pm 2m, \cdots)$ is
\begin{equation}
\begin{aligned}
C^{\operatorname{int-1D},n_{\textrm{shift}}}_{m}(k) &= |W(k)|^2\sum_l \left(\frac{k}{k-2(ml+n_{\textrm{shift}})k_N}\right)^{2p}\\
%&=\left\{
%\begin{aligned}
%&\frac{\sin^2\left(\frac{\pi k}{2k_N}\right)}{m^2\sin^2\left(\frac{\pi k}{m2k_N}-\frac{\pi n_{\textrm{shift}}}{m}\right)}\ , &\text{NGP}\ ,\\
%&\frac{\sin^4\left(\frac{\pi k}{2k_N}\right)}{m^4\sin^4\left(\frac{\pi k}{m2k_N}-\frac{\pi n_{\textrm{shift}}}{m}\right)}\left[1-\frac{2}{3}\sin^2\left(\frac{\pi k}{m2k_N}-\frac{\pi n_{\textrm{shift}}}{m}\right)\right]\ , &\text{CIC}\ ,\\
%&\frac{\sin^6\left(\frac{\pi k}{2k_N}\right)}{m^6\sin^6\left(\frac{\pi k}{m2k_N}-\frac{\pi n_{\textrm{shift}}}{m}\right)}\left[1-\sin^2\left(\frac{\pi k}{m2k_N}-\frac{\pi n_{\textrm{shift}}}{m}\right)+\frac{2}{15}\sin^4\left(\frac{\pi k}{m2k_N}-\frac{\pi n_{\textrm{shift}}}{m}\right)\right]\ , &\text{TSC}\ ,\\
%&\frac{\sin^8\left(\frac{\pi k}{2k_N}\right)}{m^8\sin^8\left(\frac{\pi k}{m2k_N}-\frac{\pi n_{\textrm{shift}}}{m}\right)}\left[1-\frac{4}{3}\sin^2\left(\frac{\pi k}{m2k_N}-\frac{\pi n_{\textrm{shift}}}{m}\right)+\frac{2}{5}\sin^4\left(\frac{\pi k}{m2k_N}-\frac{\pi n_{\textrm{shift}}}{m}\right)-\frac{4}{315}\sin^6\left(\frac{\pi k}{m2k_N}-\frac{\pi n_{\textrm{shift}}}{m}\right)\right]\ , &\text{PSC}\ .\\
%\end{aligned}
%\right.\\
&=\frac{\sin^{2p}\left(\frac{\pi k}{2k_N}\right)}{m^{2p}\sin^{2p}\left(\frac{\pi k}{2mk_N}-\frac{\pi n_{\textrm{shift}}}{m}\right)}S_{2p}\left[\sin^2\left(\frac{\pi k}{2mk_N}-\frac{\pi n_{\textrm{shift}}}{m}\right)\right]\ .
\end{aligned}
\label{eqn:shotnoise_shift}
\end{equation}
Finally, the exact shot noise in this case is
\begin{equation}
\begin{aligned}
C^{\interlace}_m(\bfk) 
%&= W^2(\bfk)\frac{\sum_{\bfn}\theta_\bfn|W(\bfk+2k_N\bfn)|^2}{W^2(\bfk)}\\
&= \sum_{(n'_x,n'_y,n'_z)\in G}C^{\operatorname{int-1D},n'_x}_m(k_x)C^{\operatorname{int-1D},n'_y}_m(k_y)C^{\operatorname{int-1D},n'_z}_m(k_z)\ .
\end{aligned}
\label{eqn:interlace_m_shotnoise_pypower}
\end{equation}

Note that group $G$ has 4 elements for $m=2$,
i.e. $(n'_x,n'_y,n'_z)=(0,0,0)$, $(0,1,1)$, $(1,0,1)$ and $(1,1,0)$.
This corresponds to the four terms in Eq.~(\ref{eqn:interlacedshotnoise_brief}).

\subsection{Bisection interlacing scheme}

Obviously in Figure~\ref{fig:Interlacing_scheme}, the equal-spacing interlacing results in anisotropic residual image distribution for $m\ge 3$.
In contrast, the bisection interlacing scheme prefers to eliminate the near images first.
To eliminate the main aliased images with $|\bfn|=1$, two interlaced grids are adopted and as a byproduct $|\bfn|=\sqrt{3}$ images are also removed.
To remove all $|\bfn|=1$ and $\sqrt{2}$ images, four interlaced grids are essential,
but $|\bfn|=\sqrt{3}$ images will be left.
To remove all $|\bfn|<2$ images, we have to increase the number of interlaced grids to 8.
We caution that it is not cost-effective to remove only one more aliased image by doubling the computation cost from 4 interlaced grids to 8.

%This corresponds to keeping only even images (from the no interlacing case) for all Cartesian directions.
%Recursively, using 16 interlaced grids only keeps the even images in each direction on the top of the residual images from 2 interlaced grids.  
We list the $\theta_\bfn$ for using $m=1$ (i.e., no interlacing), $2$, $4$, $8$ and $16$ interlaced grids,
%While conducting the higher-order interlacing scheme as described in \cite{Florian23} which the shot noise term varies with, the difference between schemes lies in $\theta_\bfn$. 
%The difference between the two schemes lies in $\theta_\bfn$.
%Given illustration samples with two-dimension are characterized by the $\theta_\bfn$ as follow:
%\begin{equation}
%\begin{aligned}
%\theta_\bfn &=\left\{
%\begin{aligned}
%\ &\frac12\left[1+\X\Y\right]\, &\text{2 interlaced}\ ,\\
%\ &\frac14\left[1+\X\right]\left[1+\Y\right] , &\text{4 interlaced}\ ,\\
%&\frac18\left[1+\Xj{1}\Yj{1}\right]\left[1+\X\right]\left[1+\Y\right]\ , &\text{8 interlaced}\ ,\\
%\end{aligned}
%\right.
%\end{aligned}
%\label{eqn:interlace_index_Florian}
%\end{equation}\par
%Which one may extrapolate into three-dimension version:
\begin{equation}
\begin{aligned}
\theta_\bfn &=\left\{
\begin{aligned}
\ &1, &\text{no interlace}\ ,\\
\ &\frac12\left[1+\X\Y\Z\right]\ , &\text{2 interlaced}\ ,\\
&\frac14\left[1+\X\Y+\Y\Z+\Z\X\right]\ , &\text{4 interlaced}\ ,\\
\ &\frac18\left[1+\X\right]\left[1+\Y\right]\left[1+\Z\right] , &\text{8 interlaced}\ ,\\
&\frac1{16}\left[1+\Xj{}\Yj{}\Zj{}\right]\left[1+\X\right]\left[1+\Y\right]\left[1+\Z\right]\ , &\text{16 interlaced}\ ,\\
&\cdots, &\cdots\ .
\end{aligned}
\right.
\end{aligned}
\label{eqn:interlace_index_Florian_3dims}
\end{equation}
In the expansion of $\theta_n$, there are exactly $m$ terms for $m$ interlaced grids. Each term takes the form $\frac{1}{m}e^{-2\pi \mathrm{i}\frac{H_xn_x+H_yn_y+H_zn_z}{H}}$, corresponding to an interlacing grid with offsets of $(H_x, H_y, H_z)$, respectively.
The residual images with $8m$ interlaced grids have the same pattern with $m$ interlaced case but the distance between residual images is scaled by a factor of two.
Generalize to cases with $m=2^{3h}$, $2^{3h+1}$ or $2^{3h+2}$ ($h=0, 1, \cdots$) interlaced grids,
%\begin{equation}
%\theta_\bfn =\left\{
%\begin{aligned}
%\ &\prodjom\frac18\left[1+\Xj{j-1}\right]\left[1+\Yj{j-1}\right]\left[1+\Zj{j-1}\right]\ , &\text{$2^{3m}$ interlaced}\ ,\\
%\ &\frac12\left[1+\Xj{m}\Yj{m}\Zj{m}\right]\prodjom\frac18\left[1+\Xj{j-1}\right]\left[1+\Yj{j-1}\right]\left[1+\Zj{j-1}\right]\ , &\text{$2^{3m+1}$ interlaced}\ ,\\
%\ &\frac14\left[1+\Xj{m}\Yj{m}+\Yj{m}\Zj{m}+\Zj{m}\Xj{m}\right]*\\ &\prodjom\frac18\left[1+\Xj{j-1}\right]\left[1+\Yj{j-1}\right]\left[1+\Zj{j-1}\right]\ , &\text{$2^{3m+2}$ interlaced}\ ,\\
%\end{aligned}
%\right.
%\label{eqn:interlace_index_Florian_3dims_n}
%\end{equation}
\begin{equation}
\begin{aligned}
\theta_\bfn 
%&=\left\{
%\begin{aligned}
%\ &\prodjom\frac18\left[1+\Xj{j-1}\right]\left[1+\Yj{j-1}\right]\left[1+\Zj{j-1}\right]\ ,\\
%\ &\frac12\left[1+\Xj{m}\Yj{m}\Zj{m}\right]\prodjom\frac18\left[1+\Xj{j-1}\right]\left[1+\Yj{j-1}\right]\left[1+\Zj{j-1}\right]\ ,\\
%\ &\frac14\left[1+\Xj{m}\Yj{m}+\Yj{m}\Zj{m}+\Zj{m}\Xj{m}\right]*\\ &\prodjom\frac18\left[1+\Xj{j-1}\right]\left[1+\Yj{j-1}\right]\left[1+\Zj{j-1}\right]\ ,\\
%\end{aligned}
%\right.\\
%%%%%%%%%%%%%%%%%%%% separation
&=\left\{
\begin{aligned}
\ &\prod_{i=x,y,z}\prodjoh\frac12\left[1+\Ij{j-1}\right]\ , &\text{$2^{3h}$ interlaced}\ ,\\
\ &\frac12\left[1+\prod_{i=x,y,z}\Ij{h}\right]\prod_{i=x,y,z}\prodjoh\frac12\left[1+\Ij{j-1}\right]\ , &\text{$2^{3h+1}$ interlaced}\ ,\\
\ &\frac14\left[1+\Xj{h}\Yj{h}+\text{2 perms.}\right]\prod_{i=x,y,z}\prodjoh\frac12\left[1+\Ij{j-1}\right]\ , &\text{$2^{3h+2}$ interlaced}\ .
\end{aligned}
\right.
\end{aligned}
\label{eqn:interlace_index_Florian_3dims_n}
\end{equation}

%One may find in two-dimensional case polynomials in the form of $1+(xy)^{\frac{1}{2^m}}$ and$(1+x^{\frac{1}{2^m}})(1+y^{\frac{1}{2^m}})$ suffice to occupy the $2^m$ space entirely, leaving one $2^mth$ the original aliasing terms. Same thing pertain in the 3-dimensional cases where we introduce polynomials in the form of $1+(xy)^{\frac{1}{2^m}}+(yz)^{\frac{1}{2^m}}+(zx)^{\frac{1}{2^m}}$
%, which is the average of four sampling at the four vertices of an equilateral tetrahedron in the 4 interlaced case. That is, four grids shift $(h/2,h/2,0)$, $(h/2,0,h/2)$, $(0,h/2,h/2)$ individually. In that case one can further deduct shot noise term analytically. 

The derivation of shot noise for $m=2^{3h}$ interlaced grids is straightforward because the summation of each direction can be done individually,
\begin{equation}
\begin{aligned}
C^\interlace_{2^{3h}}(\bfk) &= W^2(\bfk)\frac{\sum_{\bfn}\theta_\bfn|W(\bfk+2k_N\bfn)|^2}{W^2(\bfk)}\\\
&= W^2(\bfk)\prod_{i=x,y,z}\sum_{\bfn}\prodjh\frac12\left[1+\Ij{j-1}\right]\left(\frac{k_i}{k_i-2n_ik_N}\right)^{2p}\\\
&=\prod_{i=x,y,z}C^{\operatorname{int-1D}}_{2^{h}}(k_i)\ .
\end{aligned}
\label{eqn:interlace_shotnoise_3m}
\end{equation}
It is simply the multiplication of three one dimensional shot noise expressions derived in Eq.~(\ref{eqn:interlace_m_shotnoise}) for $2^h$ interlaced grids.

In case of $m=2^{3h+1}$ interlacing, following the same strategy conducted in deriving Eq.\,(\ref{eqn:interlacedshotnoise_brief}), we can obtain the summation separately and individually. 
In one dimension, the residual images with $m=2^h$ interlaced grids are divided into odd terms and even terms, with $C_m^{\interlace}=C_m^{\operatorname{int-1D,odd}}+C_m^{\operatorname{int-1D,even}}$.
Note that the even terms of $m=2^{(h-1)}$ interlacing are just the whole terms of $m=2^h$.
Thus,
\begin{equation}
C^{\operatorname{int-1D,even}}_{2^h}(k) = C^{\operatorname{int-1D}}_{2^{h+1}}(k)\ .
\label{eqn:shotnoise_even_2m}
\end{equation}
For the odd terms, they have a shift with the even terms.  Thus, 
\begin{equation}
\begin{aligned}
%\theta_{n_i}^{\textrm{odd}} &= \frac12\left[1-\Ij{m-1}\right]\prod_{j=0}^{m-1}\frac12\left[1+\Ij{j-1}\right]\\\
C^{\operatorname{int-1D,odd}}_{m=2^h}(k) &= |W(k)|^2\sum_n \left(\frac{k}{k-(2^{h+2}n+2^{h+1})k_N}\right)^{2p}\\
%&=\left\{
%\begin{aligned}
%&\frac{\sin^2\left(\frac{\pi k}{2k_N}\right)}{m^2\cos^2\left(\frac{\pi k}{m2k_N}\right)}\ , &\text{NGP}\ ,\\
%&\frac{\sin^4\left(\frac{\pi k}{2k_N}\right)}{m^4\cos^4\left(\frac{\pi k}{m2k_N}\right)}\left[1-\frac{2}{3}\cos^2\left(\frac{\pi k}{m2k_N}\right)\right]\ , &\text{CIC}\ ,\\
%&\frac{\sin^6\left(\frac{\pi k}{2k_N}\right)}{m^6\cos^6\left(\frac{\pi k}{m2k_N}\right)}\left[1-\cos^2\left(\frac{\pi k}{m2k_N}\right)+\frac{2}{15}\cos^4\left(\frac{\pi k}{m2k_N}\right)\right]\ , &\text{TSC}\ ,\\
%&\frac{\sin^8\left(\frac{\pi k}{2k_N}\right)}{m^8\cos^8\left(\frac{\pi k}{m2k_N}\right)}\left[1-\frac{4}{3}\cos^2\left(\frac{\pi k}{m2k_N}\right)+\f.rac{2}{5}\cos^4\left(\frac{\pi k}{m2k_N}\right)-\frac{4}{315}\cos^6\left(\frac{\pi k}{m2k_N}\right)\right]\ , &\text{PSC}\ .\\
%\end{aligned}
%\right.\\
&=\frac{\sin^{2p}\left(\frac{\pi k}{2k_N}\right)}{{(2m)}^{2p}\cos^{2p}\left(\frac{\pi k}{4mk_N}\right)}
S_{2p}\left[\cos^2\left(\frac{\pi k}{4mk_N}\right)\right]\ .
\end{aligned}
\label{eqn:shotnoise_odd_2m}
\end{equation}
The odd terms can also be derived by considering the fact that the whole terms of $m$ interlacing are same as the even terms of $m-1$ interlacing.
Thus,
\begin{equation}
C^{\operatorname{int-1D,odd}}_{2^{h}}(k)=C^{\operatorname{int-1D}}_{2^{h}}(k)-C^{\operatorname{int-1D,even}}_{2^{h}}(k)
=C^{\operatorname{int-1D}}_{2^{h}}(k)-C^{\operatorname{int-1D}}_{2^{h+1}}(k)
\ .
\label{eqn:interlacedshotnoise_eorelation}
\end{equation}
As such, the shot noise for $m=2^{3h+1}$ interlaced case can be written as
\begin{equation}
C^\interlace_{2^{3h+1}}(\bfk)=\prod_{i=x,y,z}C^{\operatorname{int-1D,even}}_{2^{h}}(k_i)+\sum_{(i,j,l)\in A_3} C^{\operatorname{int-1D,even}}_{2^{h}}(k_i)C^{\operatorname{int-1D,odd}}_{2^{h}}(k_j)C^{\operatorname{int-1D,odd}}_{2^{h}}(k_l)\ ,
\label{eqn:interlacedshotnoise_3m1}
\end{equation}
which is consistent with Eq.\,(\ref{eqn:interlacedshotnoise_brief}) when $h=0$. 

In case of $m=2^{3h+2}$, the summation reduces to two terms, i.e. all $n_i$'s are even and all $n_i$'s are odd, as shown in Figure~\ref{fig:Interlacing_scheme_unit}.
Therefore, the expression reads
\begin{equation}
\begin{aligned}
C^\interlace_{2^{3h+2}}(\bfk)%&=\prod_{i=x,y,z}C^{\interlace-\textrm{1D}}_{2^{m}}(k_i)-\sum_{ijl\in A_3} C^{\interlace-\textrm{1D}}_{2^{m+1}}(k_i)C^{\interlace-\textrm{1D}}_{2^{m+1}}(k_j)C^{\interlace-\textrm{1D,odd}}_{2^{m+1}}(k_l)-\sum_{ijl\in A_3} C^{\interlace-\textrm{1D}}_{2^{m+1}}(k_i)C^{\interlace-\textrm{1D,odd}}_{2^{m+1}}(k_j)C^{\interlace-\textrm{1D,odd}}_{2^{m+1}}(k_l)\\
&=\prod_{i=x,y,z}C^{\operatorname{int-1D,even}}_{2^{h}}(k_i)+\prod_{i=x,y,z}C^{\operatorname{int-1D,odd}}_{2^{h}}(k_i)\ .
\end{aligned}
\label{eqn:interlacedshotnoise_3m2}
\end{equation}

%In conclusion:
%\begin{equation}
%\begin{aligned}
%C^{\interlace}_{m}(\bfk)&=\left\{
%\begin{aligned}
%\ &\prod_{i=x,y,z}C^{\operatorname{int-1D}}_{2^{h}}(k_i)\ , &\text{$m=2^{h}$ interlaced}\ ,\\
%\ &\prod_{i=x,y,z}C^{\operatorname{int-1D,even}}_{2^{h}}(k_i)+\sum_{(i,j,l)\in A_3} C^{\operatorname{int-1D,even}}_{2^{h}}(k_i)C^{\interlace-\textrm{1D,odd}}_{2^{h}}(k_j)C^{\operatorname{int-1D,odd}}_{2^{h}}(k_l)\ , &\text{$m=2^{3h+1}$ interlaced}\ ,\\
%&\prod_{i=x,y,z}C^{\operatorname{int-1D,even}}_{2^{h}}(k_i)+\prod_{i=x,y,z}C^{\operatorname{int-1D,odd}}_{2^{h}}(k_i)\ , &\text{$m=2^{3h+2}$ interlaced}\ ,\\
%\end{aligned}
%\right.
%\end{aligned}
%\label{eqn:interlacedshotnoise_Florian}
%\end{equation}\par

The last point worth mentioning is that $C_8^\interlace(\bfk)/W^2(\bfk)$ for 8 interlaced grids given by Eq.~(\ref{eqn:interlace_shotnoise_3m}) is exactly same as $C(\bfk)/W^2(\bfk)$ with no interlacing but doubling the resolution.
In principle, at least for NGP, given the density fields on 8 interlaced grids, the density on two times finer grid can be solved out.
This explains the above equivalence.
The computational cost is roughly same, but the interlacing technique allows lower memory usage.

Finally, we present the performance of the Eq.~(\ref{eqn:interlace_m_shotnoise_pypower}), Eq.~(\ref{eqn:interlace_shotnoise_3m}), Eq.~(\ref{eqn:interlacedshotnoise_3m1}) and Eq.~(\ref{eqn:interlacedshotnoise_3m2}) in Figure~\ref{fig:shot_show_highorder}, specifically for $m=2,3,4$ with the equal-spacing interlacing scheme, and $m=2,4,8$ with the bisection interlacing scheme.  This numerical validation confirms that the above complicate theoretical derivation is correct.